%% file: main_formatfree.tex
\documentclass[11pt]{article}%
\usepackage[margin=1in]{geometry}
\usepackage{amsfonts}
\usepackage{amssymb}
\usepackage{graphicx}
\usepackage{setspace}
\usepackage[round]{natbib}
\usepackage{amsmath}%
\usepackage{amsthm}%
\usepackage{inconsolata}
\usepackage{paralist}
\usepackage{nicefrac}
\usepackage{bm}
\usepackage{mathtools}

\usepackage{booktabs}
\usepackage{lineno}

\usepackage{bm}
\usepackage{float}
\usepackage{subcaption}
\usepackage{titlesec}

\usepackage{draftwatermark}

\SetWatermarkText{DRAFT}
\SetWatermarkScale{0.8}
\SetWatermarkColor[gray]{0.99}

\usepackage{epigraph}
\usepackage{nicefrac}

\usepackage[font={small,it}]{caption}

\usepackage{mathtools}
\usepackage{lscape}
\usepackage{subcaption}
\usepackage{suffix}
\usepackage{color}
\usepackage{setspace}
\usepackage{mathrsfs}
\usepackage[inline]{enumitem}
\usepackage{framed}
\usepackage{multirow}
\usepackage{wrapfig}

\usepackage{qtree}
\usepackage{algorithm,algorithmic}

\RequirePackage[hyphens]{url}
\usepackage{xcolor}
\usepackage{hyperref}
\definecolor{darkblue}{rgb}{0.0,0.0,0.3}
\hypersetup{colorlinks,breaklinks,linkcolor=darkblue,urlcolor=darkblue,
            anchorcolor=darkblue,citecolor=darkblue}
\usepackage{empheq}
\usepackage[most]{tcolorbox}

\setlist*[enumerate]{label=(\roman*)}

\def\boxit#1{\vbox{\hrule\hbox{\vrule\kern6pt
          \vbox{\kern6pt#1\kern6pt}\kern6pt\vrule}\hrule}}
					
\usepackage{tikz}
\usetikzlibrary{calc,shapes,bayesnet}

\input{math-commands}

\theoremstyle{slplain}

\newtheorem{theorem}{Theorem}
\newtheorem{acknowledgement}[theorem]{Acknowledgement}

\newtheorem{corollary}[theorem]{Corollary}

\newtheorem{definition}[theorem]{Definition}

\newtheorem{lemma}[theorem]{Lemma}

\newtheorem{proposition}[theorem]{Proposition}
\newtheorem{remark}[theorem]{Remark}

\numberwithin{theorem}{section}

\graphicspath{{./art/}}

\title{The Curious Problem of the Normal Inverse Mean: Robustness and Shrinkage%
\thanks{J.D. was encouraged to explore this problem by Prof. Nick Polson and a blog-post by Prof. Christian Robert \citep{robert2016curious}, from which the title of this paper is adapted. This preprint is a work in progress and may be updated; feedback and corrections are welcome.}}

\author{
  Soham Ghosh \\ 
  Department of Statistics, University of Wisconsin--Madison \\ 
  \texttt{sghosh39@wisc.edu}
  \and
  Uttaran Chatterjee \\
  H.~Milton Stewart School of Industrial and Systems Engineering, Georgia Tech \\
  \texttt{uchatterjee8@gatech.edu}
  \and
  Jyotishka Datta \\
  Department of Statistics, Virginia Tech \\
  \texttt{jyotishka@vt.edu}
}

\date{\today}

\begin{document}
\maketitle

\begin{abstract}
\noindent In astronomical observations, the estimation of distances from parallaxes is a challenging task due to the inherent measurement errors and the non-linear relationship between the parallax and the distance. This study leverages ideas from robust Bayesian inference to tackle these challenges, investigating a broad class of prior densities for estimating distances with a reduced bias and variance. Through theoretical analysis, simulation experiments, and application to data from the \textit{Gaia} Data Release~1 (GDR1), we demonstrate that heavy-tailed priors provide more reliable distance estimates, particularly in the presence of large fractional parallax errors. Theoretical results highlight the ``curse of a single observation,'' where the likelihood dominates the posterior, limiting the impact of the prior. Nevertheless, heavy-tailed priors can delay the explosion of posterior risk, offering a more robust framework for distance estimation. The findings suggest that reciprocal-invariant priors with polynomially decaying tails, such as the Half-Cauchy and Product Half-Cauchy, are particularly well suited for this task, providing a balance between bias reduction and variance control.
\end{abstract}


\section{Introduction}
A parallax refers to the apparent shift in the position of an object when observed from two different vantage points. In the context of stellar astronomy, the annual parallax is the most common form, which is the apparent displacement of a star observed from Earth at opposite points in its orbit around the Sun. \citet{parallaxhistory} pens down the rich history of studying stellar parallaxes in astronomy.  With the advent of modern astronomy, technological advancements have led to highly precise measurements of parallaxes, particularly through space-based observatories like the Hipparcos and Gaia missions. Accurate distance measurements are crucial for determining the intrinsic properties of stars, such as their luminosity, mass, and size. In this endeavor,  parallax measurements provide the most direct and reliable method for determining the distances to stars close to Earth. \citet{reidsurvey} revisits the first stellar parallaxes to estimate distances. The underlying principle involves measuring the apparent motion of a star against a distant background as the Earth orbits the Sun. By observing a star at opposite points in Earth's orbit (six months apart), astronomers can measure the angle of apparent shift in the star's position. The distance to a star (in parsecs) given by $r$ can be calculated as the reciprocal of the parallax angle $\omega$ in arcseconds using the reciprocal relation $r = 1/\omega$. 

\begin{figure}[!ht]
     \centering
     \includegraphics[width=0.8\linewidth]{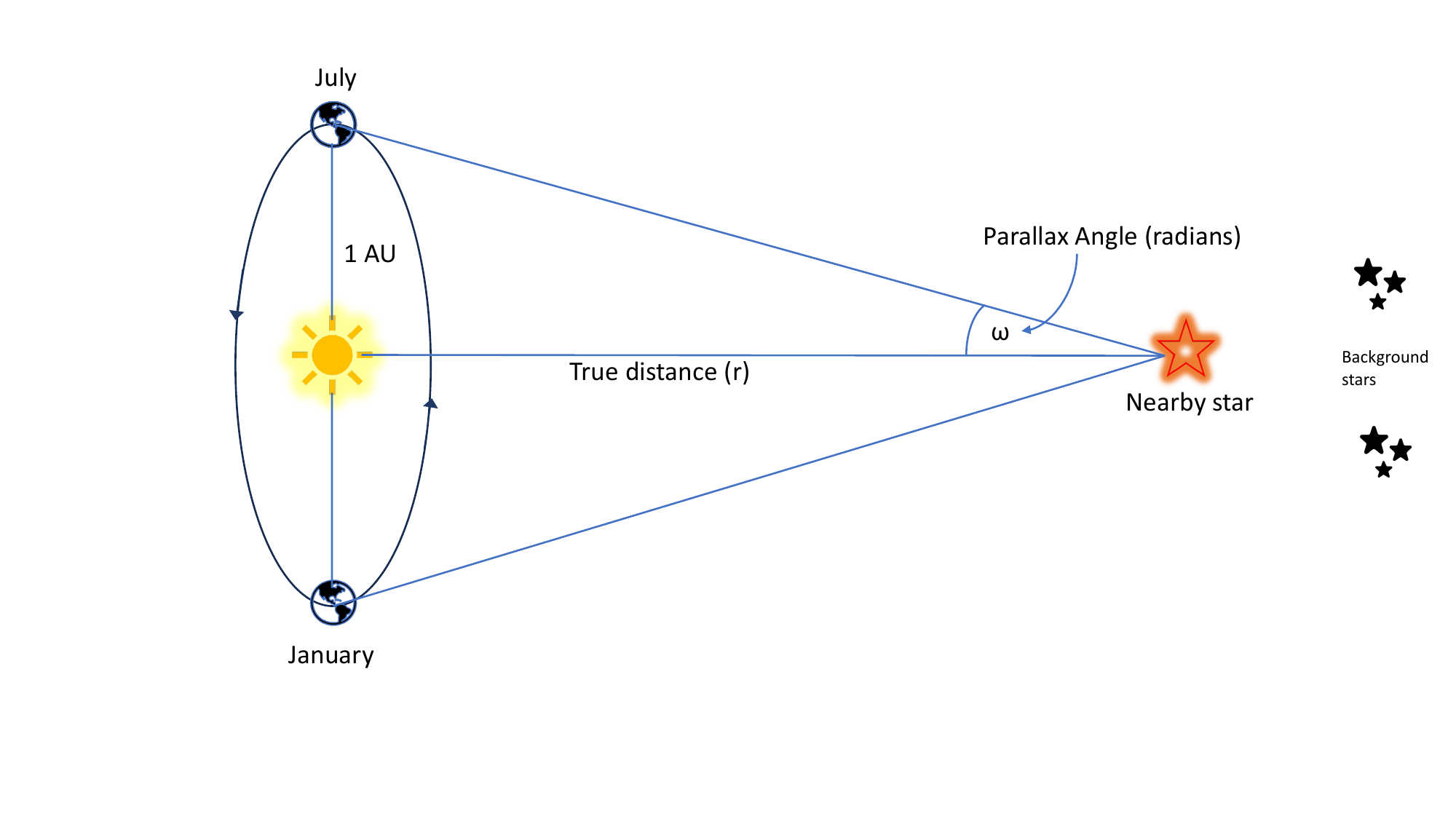}
     \caption{\footnotesize{Trigonometric estimation of distances from stellar parallaxes}}
     \label{fig:parallaxmeasure}
\end{figure}

This relationship can be best understood through basic trigonometry, if we take the Earth-Sun distance (1 Astronomical Unit) as the baseline for the triangle in Figure \ref{fig:parallaxmeasure}, the distance to the star and the parallax angle are related as $\tan \omega = {1 \ \text{AU}}/{r}.$ For small angles $\omega$, which typically occur even for stars not too far from Earth, $\tan \omega \approx \omega$ when $\omega$ is measured in radians. The approximate relation then becomes: $ r \approx {1 \ \text{AU}}/{\omega}.$
When $\omega$ is measured in arcseconds, $r$ is given in parsecs (pc), where 1 parsec corresponds to the distance at which a star would have a parallax angle of 1 arcsecond.

Although we get a simple relation between the two, inverting a parallax to
give a distance is only appropriate when we have no measurement errors, which is an unrealistic assumption to make in practical use-cases. \citet{Bailer_Jones_2015} argues that estimating distances from parallaxes is not trivial, especially when the fractional parallax error (defined as $f=\sigma_\omega/\omega$ as in \eqref{model}) is large (greater than 20\%). Thus, determining the distance from a parallax measurement is translated into an inference problem, which in its simplest version is estimating the distance $r$ (measured typically in parsec, or `pc'), given the parallax $\omega$ (measured in arcseconds) and its uncertainty $\sigma_{\omega}$.

We consider the following (simplified) model, as presented in existing literature on parallax estimation \citep{Bailer_Jones_2015,Astraatmadja_2016}. Denoting the observed parallax as $\omega$, a noisy measurement corresponding to the true distance $r$, the model is: 
\begin{equation}\label{model}
    \omega= \frac{1}{r}+\epsilon_\omega; \quad \epsilon_\omega \sim N(0, \sigma_\omega^2),
\end{equation}
where $\sigma_\omega^2$ quantifies the error in measuring the parallax $\omega$. Hence, $\sigma_\omega ^2$ can be either a function of $\omega$ known up to a functional form (or, as discussed in \citet{debruijne12} can be estimated from another astronomic measurement model) or an unknown function of $\omega$. We will address both cases later in detail in Section \ref{sec:method}.  Hence, the likelihood of the our model in \eqref{model} happens to be: 
\begin{gather}
        \omega \mid r,\sigma_{\omega} \sim \NormRV\left(\frac{1}{r}, \sigma_{\omega}^2\right) \Rightarrow 
        P(\omega \mid r,\sigma_{\omega}) = \frac{1}{\sqrt{2 \pi} \sigma_{\omega}} \exp{\left[-\frac{1}{2\sigma_{\omega}^2}\left(\omega-\frac{1}{r}\right)^2\right]} \nonumber \\
        \L(r ; \omega) \propto \exp{\left[-\frac{1}{2\sigma_{\omega}^2}\left(\frac{1}{r^2} - 2 \frac{\omega}{r} \right)\right]}\label{eq:likelihood}. 
\end{gather}
Given the parallax $\omega$ and its uncertainty $\sigma_{\omega}$, the statistical challenge is to derive an accurate estimate of the true distance $r$. Although this is rooted to one of the simplest and the most well-studied models namely, the Normal model, the problem is far from trivial! The complications in inference of the distance $r$ arises due to the following two primary reasons: firstly, we have a single parallax measurement. This limitation arises due to various factors such as the limited operational lifetimes of observatories, the faintness of distant stars making repeated measurements challenging, or the substantial time and resource commitments required for multiple observations. The second challenge is induced by the inverse (nonlinear) relationship between parallax and distance:  even a symmetrical error in parallax will lead to an asymmetrical error in the derived distance. 

\begin{remark}{Broader Applications:}
While the primary motivation for this study is astronomical distance estimation using parallax measurements, the statistical problem of inferring the reciprocal of a normal mean arises in several other domains, as noted by \cite{Withers2013}. In \textbf{nuclear physics}, it arises in determining the momentum of charged particles from track curvature \citep{Lamanna1981, Treadwell1982}. In \textbf{econometrics}, it appears in structural equation models and economic multipliers \citep{Zellner1978, Zaman1981a, Braulke1982}. Applications also include \textbf{medical imaging}, where it aids in detecting rolling leukocytes \citep{Dong2005, Sahoo2006}, and \textbf{finance}, where it underlies risk-adjusted performance measures such as the Sharpe ratio \citep{Powers2009, Kaluszka2003}. Further instances occur in \textbf{signal processing} \citep{Brown2001} and \textbf{fluid dynamics} \citep{Frommelt2008}. 
\end{remark}

\begin{figure}[h!]
  \includegraphics[width=\linewidth]{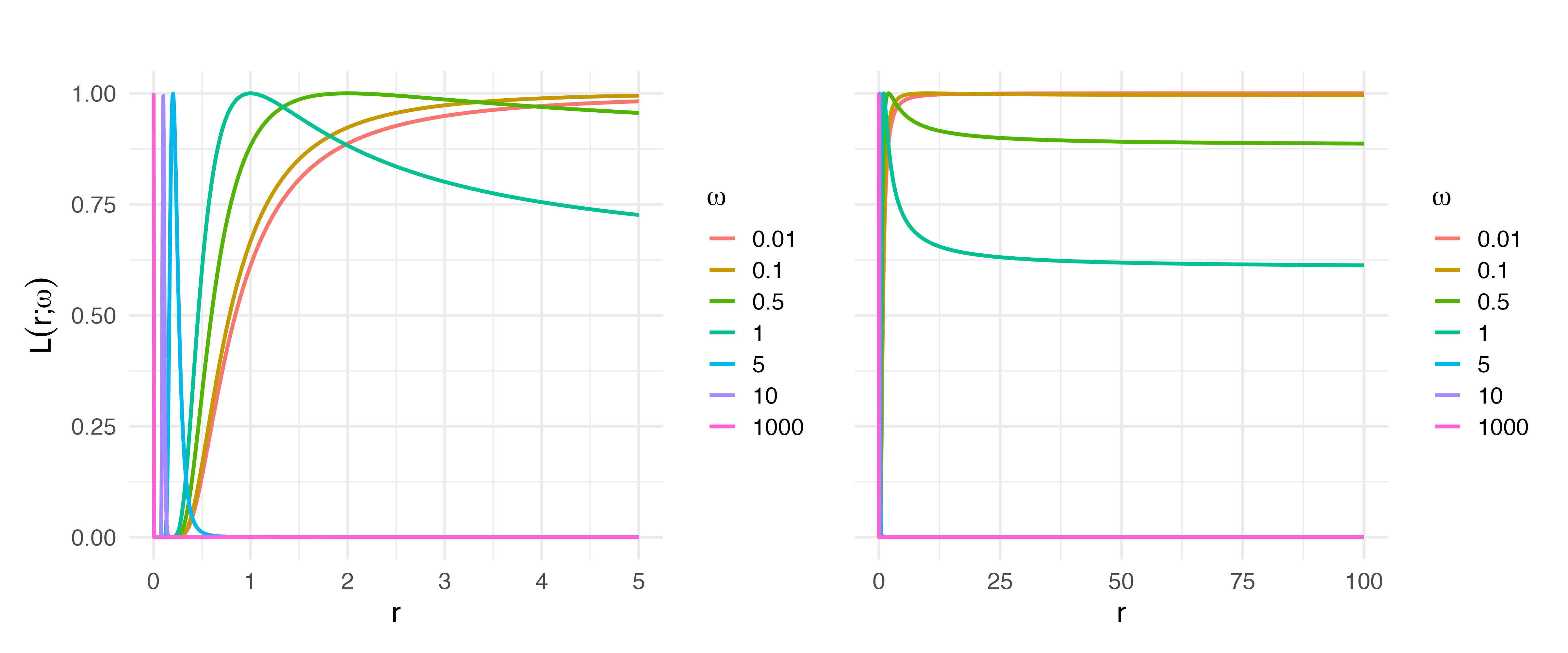}
   \caption{\footnotesize{Plot of the likelihood as a function of $r$ for $\sigma_{\omega} = 1$ to show the effect of $\omega$ on the tail-behaviour. The left panel shows the likelihood's shape for nearby distances. The right panel shows that for small $\omega$, the likelihood becomes very broad and assigns significant probability to very large distances.}}
   \label{fig:fig1}
\end{figure}

As an initial exploration, we plot the likelihood in \eqref{eq:likelihood} as a function of $r$ for a fixed value of $\sigma_{\omega}=1$ to exhibit the effect of observed $\omega$ on the shape and the tails of the likelihood (see Figure \ref{fig:fig1}). As noted from Figure \ref{fig:fig1}, for large $\omega$ values (e.g., 10, 1000), the likelihood tails off quickly as $r$ increases, indicating that very large distances become highly improbable. The peaks of the likelihood function become more distinct and sharper around $r={1}/{\omega}$. For smaller $\omega$ values (e.g., 0.01, 0.1), contrary to a rapid decline, the likelihood exhibits a pronounced tail extending towards larger $r$. This indicates that even though the most likely estimate of $r$ is near ${1}/{\omega}$, there is a significant probability attributed to larger distances. This tail behaviour can be attributed to the fact that the transformation from parallax to distance $r={1}/{\omega}$ is non-linear, complicating the propagation of errors. For small parallaxes (i.e., large distances), the distribution of ${1}/{\omega}$ becomes highly skewed and long-tailed. This skewness and heavy tails in the distribution of estimated distances indicate that standard statistical methods, which often assume normality or symmetric distributions, may not perform well. Thus, it is easy to see that the na\"ive approach of constructing a credible region with ${1}/{\omega} \pm {\sigma_{\omega}}/{\omega^2}$ using the Delta method is unreliable and noisy, especially near the origin \citep{Bailer_Jones_2015}. 
\\


It also follows from \eqref{eq:likelihood} that the Fisher's information for $r$ is given by:
\begin{equation}\label{eq:jeffreys}
  I(r) = -\mathbb{E}\left[\frac{\partial^2 \log \L(r; \omega)}{\partial r^2}\right] = -\mathbb{E}\left[\frac{1}{\sigma_{\omega}^2}\left(\frac{2\omega}{r^3}- \frac{3}{r^4}\right)\right] = \frac{1}{\sigma_{\omega}^{2} r^{4}}  
\end{equation}
The Fisher's information \eqref{eq:jeffreys} is unbounded near $r = 0$, which in turn indicates high sensitivity of inference near the origin, i.e., small changes in the parameter around $0$ result in large changes in the likelihood. This could indicate that the model is extremely sensitive to the value of $r$ when it is close to $0$. 

Estimating the reciprocals and ratios of means and regression coefficients have been investigated thoroughly in the frequentist literature \citep[e.g.][]{ZELLNER1978127,Kuhetal}. For example, the Minimum Expected Loss (MELO) estimators have been proposed specifically for estimating the reciprocal means of Normal populations \citep{zellnerMELO}. They offer promising alternatives as estimators in problems where the maximum likelihood estimators do not possess finite moments and have
infinite risk relative to quadratic and many other loss
functions. In our setting, we can derive the MELO estimator of the distance $r$ under the quadratic loss as ${\omega}/{(\omega^2 + \sigma_{\omega} ^2)}$, whereas the MLE is given by ${1}/{\omega}$. Note that although the MELO estimator attempts to mitigate some of the issues with direct reciprocal estimation by incorporating $\sigma_{\omega} ^2$ in the denominator attempting to stabilize the variance, it does not capture the nuance of the reciprocal relationship between the distance and the parallax. This is because as the measured parallax $\omega$ approaches zero, the MELO estimator of the true distance also approaches zero, which is the exact opposite of what is expected, as a smaller parallax implies a larger distance.

Given the challenges with both the MLE and MELO estimators, a Bayesian approach offers a conceptual advantage by incorporating prior knowledge about the distribution of $r$. However, since we have a single observation, it is not favorable in terms of checking whether the prior is at all correctly specified, which can lead to biased or misleading estimates. However, there exists a class of default priors, named Single Observation Unbiased Priors (SOUP) which can be obtained if the corresponding posterior mean of the parameter based on a single observation is an unbiased estimator of the parameter \citep{SOUP}. By ensuring that the posterior distribution is unbiased with just a single observation, SOUP priors guard against the skewness induced by poorly chosen priors by aligning the posterior more closely with the actual data characteristics. Unfortunately, it has been well established that SOUP priors do not exist for non-linear transformations of Normal mean parameters \citep{SOUP,Hartigan98}.  

From an objective Bayesian perspective, our main question is: {\em ``How do we build a suitable prior for such a likelihood shape?"} \citet{Bailer_Jones_2015} compared different priors and shows that the improper uniform prior or the even the prior with a sharp cut-off at some distance give very poor distance estimates (large bias and variance), and prescribed using a prior that asymptotically decays to zero at infinity, based on empirical evidences. More recently, \citet{Bailer_Jones_2021} presented a methodology for estimating distances to stars using data from \textit{Gaia} Early Data Release 3. The study introduced geometric and photo-geometric distance estimates, utilizing a new three-parameter distance prior that replaces the simpler exponential model used in previous works. This approach enhances the precision of distance estimates across the sky by using advanced statistical models that better accommodate the complex spatial distribution of stars within our galaxy. However, it was still not robust enough to handle large fractional parallax errors. In sections \ref{sec:theory} and \ref{sec:simuls}, we demonstrate that heavy-tailed priors with polynomial tail decay rates indeed work better for this problem compared to priors with exponentially decaying tails. This aligns with the philosophy articulated in \cite{gelman2006prior}, which posits that ``a prior can in general only be interpreted in the context of the likelihood with which it will be paired." The suitability of heavy-tailed priors in our context is also supported by seminal papers in Bayesian robustness literature, such as \cite{dawid1973posterior}, who argued that the prior ``should not wag its tail too vigorously" in order to effectively reject outliers for location families. 

We turn to the notions of credence \citep{o1990outliers} and p-credence \citep{p-cred07}, which helps to characterize overall ``thickness" of a prior distribution's tails beyond just the extreme ends. Our theoretical exploration leads to several key insights including the fact that, the likelihood, particularly in the case of a single observation, tends to dominate the posterior, overshadowing the influence of the prior. This phenomenon is encapsulated in our discussion detailed in Section \ref{sec:theory} as the ``curse of a single observation," where we highlight how heavy-tailed priors, despite their robust properties, may not substantially alter the posterior relative to the likelihood.

\noindent \textbf{Organization of the paper:} The paper is segmented as follows. A short history of heavy-tailed priors in Bayesian literature and the ideas of regular variation and credence are described in Section \ref{sec:rv}. Existing approaches, as elucidated in \citet{Bailer_Jones_2015,Astraatmadja_2016}, are briefly reviewed and remarked upon in Section \ref{sec:existing}. Our proposed priors with rationale behind each of the candidates is in Section \ref{sec:method}. 
Our theoretical results show that the p-credence of the posterior can be broadly classified on the basis of the tail-decay rates of $r$ with $\nicefrac{1}{\omega}$, as shown in Section \ref{sec:theory}. Then, in Section \ref{sec:risk}, we provide posterior risk behaviors for different loss functions and a wide variety of non-degenerate prior distributions with polynomial and exponential tails, as broadly classified by the tail-decay rate under p-credence. We demonstrate the clear advantage of using polynomial-type decay priors over exponential-type decay priors through two elaborate simulation experiments in Section \ref{sec:simuls}. The \textit{Gaia} Data Release 1 (GDR1) dataset is analyzed in Section \ref{sec:realdata} by contrasting their naive inverse parallax estimates against our posterior distance estimates. We conclude the paper with a brief discussion in Section \ref{sec:disc}. The proofs of some of our theoretical results along with MCMC diagnostics for the simulation results and posterior predictive checks are specified in the Appendix.

\subsubsection{Useful Notations:} For any functions $f,g:\mathbb{R}\to \mathbb{R}$ we write;
\begin{itemize}
\item $f=o(g)$ if there $f(x)/g(x)\to 0$ as $x \to \infty$.
\item $f=\mathcal{O}(g)$ if there exist a constant $K\ge 0$ such that $|f(x)|\le K g(x)$ or all $x\ge x'$ for some $x'\in \mathbb{R}$. 
\item $f\sim g$ if $f(x)/g(x)\to 1$ as $x \to \infty$. 
\item $f \asymp g$ if there exist positive constants $c_1$ and $c_2$ such that $c_1 g(x) \le f(x) \le c_2 g(x).$
\item $f=\Omega(g)$ if there exists $k\ge 0$, such that $kg(x)\le |f(x)|$ for $x\ge x''$ for some $x''\in \mathbb{R}$.
\item $f=\Theta(g)$ if $f=\mathcal{O}(g)$ and $f=\Omega(g)$.
\end{itemize}
We denote any measurable set $A \subseteq\mathbb{R}$ we denote the indicator function of the set $A$ by $\mathbb{I}(A)$. 

For any random observation $x$, drawn from a population described by a parametric distribution with density function $f_\theta$ where $\theta$  is a parameter in some well-defined parameter space $\Theta$, we denote the likelihood function for the parameter $\theta$ by $\L(\theta ; x)$.

\section{Heavy-tailed priors: Credence}\label{sec:rv} 

Our choice of prior builds upon the long and rich history of heavy-tailed distributions in probability theory and Bayesian robustness modeling to resolve conflicts between the prior and likelihood, going back to \citet{de1961bayesian, lindley1968choice}, and \citet{dawid1973marginalization}, and more recently \citet{o1979outliers,o1990outliers, andrade2006bayesian,andrade2011bayesian, o2012bayesian}. Various properties of these heavy-tailed priors were utilized in designing continuous shrinkage priors for high-dimensional parameter space. Such shrinkage priors provide two simultaneous benefits: first, a spike at zero that helps squelching small, noisy parameters closer to zero and the thick tails for robustness in the tails. The latter is ensured by a result by \cite{barndorff1982normal} who showed that for any mixing distribution with slowly-varying tail behavior, the corresponding normal variance mixture will also have slowly varying tails. These are the so-called `global-local' shrinkage priors \citep{bhadra2019lasso}, with the most popular method being the `horseshoe' prior \citep{carvalho2010horseshoe}, so named for the induced ${\rm Beta}(1/2, 1/2)$ prior on the shrinkage factors. The horseshoe prior and its many variants have been immensely popular in the Bayesian literature for the last 15 years and have attained `state-of-the-art' status for high-dimensional data with low-dimensional structures \citep[see][]{bhadra2019lasso}.

There are several ways of characterizing the tail-decay of a prior distribution, e.g. polynomial decay, regular variation \citep{mikosch1999regular, barndorff1982normal}, credence \citep{o1979outliers} and p-credence \citep{p-cred07}.  However, while regular variation tails offer a powerful tool to characterize the asymptotic behavior of distributions, they do not provide a complete picture of the thickness of the distribution's tails across its entire domain. This is where the concept of \emph{credence} becomes valuable. Credence, as defined by \citet{o1990outliers}, provides a more comprehensive measure that captures not just the tail behavior but the overall thickness of the distribution.

\begin{definition} (\cite{o1990outliers})
    A density $f$ on $\mathbb{R}$ has a credence $c$ if, there exist constants $0<k \le K < \infty$ such that for all $x \in \mathbb{R}$ we have,
    \begin{equation}
        k \le (1+x^2)^{c/2}f(x) \le K.
    \end{equation}
    We write $\text{cred}(f)=c$.
\end{definition} 

A credence of $c < \infty$ ensures that for large $|x|$, $f(x)$ is of the order $|x|^{-c}$. While the idea of credence is one of its kind in a wide class of problems where we might wish to consider distributions whose tails are similar to the tails of a Students' $t$ distributions, it encompasses only a small class of densities. In fact, the credence of the Normal distribution, which is of paramount importance to us, being the likelihood, has infinite credence. Thus, no matter the choice of our prior, whether it has finite or infinite credence, the posterior will always have an infinite credence \citep{o1990outliers}. This makes it difficult to study or compare the tail properties of the prior distribution because the posterior's tail behavior is dominated by the likelihood. In general, plain credence only distinguishes algebraic tail orders and assigns \emph{infinite} credence to many common models (e.g., the Normal), so it cannot compare priors with an exponential tail to a Gaussian likelihood, nor can it separate polynomial from logarithmic refinements. To compare likelihoods and priors on a common scale and to capture logarithmic,
polynomial, and exponential regimes within one template, we adopt an extension
of credence that calibrates densities against a flexible reference family. In that regard, \citet{p-cred07} defined the generalized exponential power distribution as follows: 
\begin{definition}(\textbf{Generalized Exponential Power Family})\label{def:gep}\\
    The class of distributions with density:
    \begin{align} \label{eq:gep}
        p(z|\gamma,\delta,\alpha,\beta) & \propto \max(|z|,z_0)^\alpha \log ^{\beta}(\max(|z|,z_0)) \nonumber\\
        & \exp \left(-\delta \max(|z|,z_0)^{\gamma} \right) 
    \end{align}
        \vspace{-0.2em}
    satisfying:
    \begin{itemize}
        \item [(1)] $z_0 > 1$, if $\beta \neq 0$; or $z_0 > 0$, if $\alpha < 0, \beta = 0$
        \item [(2)] $\alpha + \frac{\beta}{\log (z_0)} \le \delta \gamma z_0 ^{\gamma}$
        \item [(3)] $\alpha \le -1$, if $\gamma = 0$
        \item [(4)] $\beta < -1$, if $\gamma = 0$ and $\alpha = -1$
    \end{itemize}
\end{definition}

Definition \ref{def:gep} is further used to accommodate the logarithmic, polynomial, and exponential tail behaviors in the original notion of credence by further introducing \emph{p-credence}.
\begin{definition}(\cite{p-cred07})
    A density $f$ on $\mathbb{R}$ has p-credence $(\gamma, \delta, \alpha, \beta)$ if there exist constants $0<k \le K < \infty$ such that for all $x \in \mathbb{R}$ we have, 
    \begin{equation}
        k \le \frac{f(x)}{p(x \mid \gamma, \delta, \alpha, \beta)} \le K.
    \end{equation}
    Where $p(x \mid \gamma, \delta, \alpha, \beta)$ is a density from the \textit{Generalized Exponential Power} family defined in (\ref{eq:gep}). We write, \text{p-cred}$(f) = (\gamma, \delta, \alpha, \beta)$.
\end{definition}
It is to be noted that p-credence is defined for most of the usual symmetric densities supported on $\mathbb{R}$ (e.g. Normal, Student's-t, Laplace etc.). The introduction of p-credence allows for a unified framework to study a wide variety of tail behaviors -- 
    \begin{itemize}
                \item [(i)] Logarithmic tails when $\beta \neq 0$
                \item [(ii)] Polynomial tails when $\alpha \neq 0$
                \item [(iii)] Exponential tails when $\gamma \neq 0$
    \end{itemize}
However, we are particularly interested in using the notion of p-credence to study the `dominance' relation between the densities of our prior and the likelihood. The meaning of a density $f$ `dominating' another density $g$ is outlined in the following definition.
\begin{definition}
    Let $f$ and $g$ be any two densities on $\mathbb{R}$. We say that, 
    \begin{itemize}
        \item[(i)] $f$ dominates $g$ if there exists a constant $k>0$ such that, 
                \[f(x) \ge kg(x) \ \ \forall x \in \mathbb{R}:\]
                We write, $f \succeq g$.
        \item[(ii)] $f$ is equivalent $g$ if both $f \succeq g$ and $g \succeq f$. We write, $f \approx g$;
        \item[(iii)] $f$ is strictly dominates $g$ $(f \succ g)$, if $f \succeq g$ but $g \nsucceq f$.
    \end{itemize}
\end{definition}
Because p-credence represents tail thickness via the tuple $(\gamma,\delta,\alpha,\beta)$, it induces a natural (lexicographic) ordering on densities; the next result formalizes when one density dominates another in this sense.
\begin{lemma}(Proposition 1 in  \cite{p-cred07})\label{lem:1}
    Let $f$ and $g$ be two densities on $\mathbb{R}$ such that p-cred($f$) $=(\gamma, \delta, \alpha, \beta)$ and p-cred($g$) $=(\gamma',\delta',\alpha',\beta')$, then
    \begin{itemize}
        \item[(i)] $f \approx g$ if $\gamma=\gamma'$, $\delta=\delta'$, $\alpha=\alpha'$ and                    $\beta=\beta'$.
        \item[(ii)] $f \succ g$ if:
                    \begin{itemize}
                        \item[(a)] $\gamma' > \gamma$;
                        \item[(b)] $\gamma'=\gamma$,  $\delta' > \delta$;
                        \item[(c)] $\gamma'=\gamma$,  $\delta'=\delta$,  $\alpha' < \alpha$:
                        \item[(d)] $\gamma'=\gamma$,  $\delta'=\delta$,  $\alpha'=\alpha$, $\beta'              <\beta$;
                        
                    \end{itemize}
    \end{itemize}
\end{lemma}
The proof of this lemma is purely technical and is elaborated in \citet{p-cred07}. We can interpret the strict dominance ($f \succ g$) in the following manner:
\begin{itemize}
    \item[(a)] $\gamma' > \gamma$: The $\gamma$ parameter controls the exponential decay. A smaller $\gamma$ means a slower, heavier exponential tail.
        
    \item[(b)] \textbf{$\gamma' = \gamma$ and $\delta' > \delta$:} When $\gamma' = \gamma$, the densities $f$ and $g$ have similar exponential tail decay rates. However, $\delta$ influences the scale of this decay. If $\delta' > \delta$, the exponential term $\exp(-\delta' |z|^{\gamma})$ for $g$ decays more slowly than $\exp(-\delta |z|^{\gamma})$ for $f$, meaning $g$ has a heavier tail.
    
    \item[(c)] \textbf{$\gamma' = \gamma$, $\delta' = \delta$, and $\alpha' < \alpha$:} Here, $\alpha$ influences the polynomial decay of the tails. If $\gamma$ and $\delta$ are equal, but $\alpha' < \alpha$, then $g$ has a thicker tail in the polynomial sense, meaning $f$ dominates $g$ by decaying faster polynomially.
    
    \item[(d)] \textbf{$\gamma' = \gamma$, $\delta' = \delta$, $\alpha' = \alpha$, and $\beta' < \beta$:}
    Finally, if all other parameters are equal, but $\beta' < \beta$, the logarithmic decay influences the dominance relation. A lower $\beta'$ means $g$ has a slower logarithmic decay than $f$, making $f$ dominate $g$ by decaying faster logarithmically.
\end{itemize}
The conditions laid out in Lemma \ref{lem:1} provide clear guidelines for constructing priors that dominate the likelihood or be dominated by it. If robustness is desired, one might choose a prior that dominates the likelihood, ensuring that the posterior is not overly influenced by potentially outlier-prone data, which in our case translates to having small values of the parallax $\omega$.

\section{Existing Approaches}\label{sec:existing}
Traditional frequentist approaches do not tend to work in this scenario, as correctly pointed out by \citet{Bailer_Jones_2015} and \citet{Astraatmadja_2016}. From the measurement model itself, we can infer that for $1/r$ the combined intervals $[\omega-2\sigma_{\omega}, \omega]$ and $[\omega, \omega+2\sigma_{\omega}]$ contain about $95.4 \%$ of the total probability of the distribution $P(\omega|r, \sigma_{\omega})$. Inverting, for $r$, each interval
$\left[\nicefrac{1}{\omega},\nicefrac{1}{(\omega-2\sigma_{\omega})}\right]$  and $\left[\nicefrac{1}{(\omega+2\sigma_{\omega})},\nicefrac{1}{\omega}\right]$ tend to have the same coverage probability since the transformation $1/r \to r$ is monotonic and hence preserves the probability. Suppose, if we have a large fractional parallax error $f = \nicefrac{\sigma_{\omega}}{\omega} = 1/2$, the upper distance interval becomes $[{1}/{\omega},\infty)$ -- which makes the confidence interval non-informative, as some finite amount of probability in the likelihood will always correspond to an undefined distance. 

The natural inclination is to use a Bayesian approach where we infer the distance in a probabilistic sense if we adopt a prior assumption about it, independent of the parallax we have observed. \citet{Bailer_Jones_2015} compared four different priors on $r$, \textit{viz.}
\begin{description}
    \item[1. An improper uniform:] $\pi(r) = \mathbb{I}\{r>0\}$, which in turn leads to an improper posterior (see Fig. \ref{fig:fig1}). We would also like to note that an improper uniform is not considered non-informative in the strictest sense \citep[see e.g.][]{gelman2014understanding, jeffreys1972methods}. 
    \item [2. A proper uniform:] $r \sim \UnifRV(0, r_{\text{lim}})$, where $r_{\text{lim}}$ is the maximum expected distance of a star in our galaxy, e.g. $r_{\text{lim}} = 10^3$ pc. 
    \item [3. A `constant volume density' prior:] With $r_{\text{lim}}$ having the same interpretation, the constant volume density prior is given by:
    \begin{equation*}
        \pi_{cv}(r \mid \omega, \sigma_{\omega}) = \frac{r^2}{\sigma_{\omega}} \exp\left[ -\frac{1}{2\sigma_{\omega}^2}\left(\omega - \frac{1}{r}\right)^2  \right] \mathbb{I}\{0 \le r \le r_{\text{lim}} \}. \label{eq:cvd-prior}
    \end{equation*}
    This is called a `constant volume density' prior by \citet{Bailer_Jones_2015}, because it is derived from the physical assumption that: \emph{the probability of finding a star in a shell with inner and outer radii $(r, r + {\rm d}r)$ is proportional to the volume of that shell $4\pi r^2 {\rm d} r$.}
    \item [4. An exponentially decreasing (ED) volume density prior:] This is equivalent to a Gamma $(3,{1}/{L})$ prior, replaces the sharp truncation by a tapering tail, \textit{i.e.}
    \begin{equation*}
        \pi_{\text{ED}}(r \mid L) = \frac{1}{2L^3} r^2 \exp\left(-\frac{r}{L}\right) \label{eq:edvd-prior}.
    \end{equation*}
    Typically, $L$ is taken to be of the same order as $r_{\lim} = 10^3 \ \text{pc}.$
\end{description}

Beyond this point, we will use the terms Gamma and ED priors interchangeably throughout the manuscript. The superior performance of the ED prior in reducing bias and variance in estimated distances, particularly when using the posterior mode has been highlighted previously \citep{Bailer_Jones_2015,Astraatmadja_2016}. This advantage holds notably under conditions of non-positive parallaxes and inevitably, large fractional parallax errors: $f_{\text{true}} = \sigma_{\omega} r_{\text{true}} \le 0.4$, where $r_{\text{true}}$ is the true distance taken into the simulation setting. Simulation results demonstrate a stark contrast in performance among different priors: the standard deviation of the posterior mode for the improper uniform prior sharply increases as $f_{\text{true}}$ surpasses 0.2. Although the proper uniform prior shows slight improvement, maintaining stability until $f_{\text{true}}$ reaches 0.25, it still under-performs compared to the constant volume density prior, which robustly controls the standard deviation of the posterior mode up to a $f_{\text{true}}$ of 0.35. 

\section{Proposed Methodology: Heavy-tailed priors for Distance Estimation}\label{sec:method}
Following the parallax model in \citet{Bailer_Jones_2015}, $\sigma_\omega$ quantifies the measurement error incurred while measuring parallax $\omega$. For computational brevity in this section, we assume that $\sigma_{\omega}$ is known. The scenario involving unknown $\sigma_{\omega}$ has been deferred to Appendix \ref{sec:unknownsigma}, by placing an Inverse Gamma prior on $\sigma_{\omega}$. \citet{Bailer_Jones_2015}'s suggestions have been instrumental in the choice of informative priors for solving the problem of exploding standard deviations of the posterior estimators. He does not recommend the use of a prior with a sharp cutoff (for instance, the proper uniform prior), because it introduces significant biases for low-accuracy measurements at all distances. He endorses the idea of using a prior which converges asymptotically to zero as the distance becomes large enough. We back that idea by proposing the following heavy-tailed candidate priors:



\begin{enumerate}
    \item \textbf{Reciprocal-Gaussian Prior (RG$^{+}$)}: We consider the truncated Reciprocal Gaussian  prior $\text{RG}^{+}(0, \sigma^2)$
     \[ \pi(r) = \sqrt{\frac{2}{\pi}}\frac{1}{\sigma r^2}\exp \left(-\frac{1}{2\sigma^2r^2}\right), \ \ r>0. \] 
     \emph{It serves as a conjugate prior to within the measurement model used.}
    This can be observed easily while computing the posterior as follows:
\begin{align*}
    \pi(r|\omega) & \propto \exp\left[-\frac{1}{2\sigma_\omega^2} \left\{\frac{1}{r^2}-\frac{2\omega}{r}\right\}\right]\frac{1}{r^2}\exp\left(-\frac{1}{2\sigma^2r^2}\right) \\
                  & \propto \frac{1}{r^2}\exp\left[ -\frac{1}{2\tau_\omega^2}\left(\frac{1}{r}-\frac{\tau_\omega^2\omega}{\sigma_\omega^2}\right)^2\right].
\end{align*}

Where $\frac{1}{\tau_\omega^2} :=\frac{1}{\sigma_\omega^2}+\frac{1}{\sigma^2}.$ Implying that,
\[r|\omega, \sigma_\omega \sim \mathrm{RG}^+\left(\frac{\tau_\omega^2\omega}{\sigma_\omega^2},\tau_\omega^2 \right).\]

\item \textbf{Inverse-Gamma prior}: Considering a small shape parameter $\alpha>1$, it features a \textit{heavier tail} compared to the ED prior, making it suitable for applications requiring robust handling of smaller parallax measurements. The Inverse-Gamma $(\epsilon,\epsilon)$ originally proposed by \citet{Browneinvgamma}, is a pretty commonly used ``diffuse" prior in the literature, for small values of $\epsilon$-- sometimes a default prior in Bayesian modeling, particularly when dealing with variance components in hierarchical or mixed-effect models.
\item \textbf{Half-Cauchy prior}: The Jeffreys prior, based on the square root of Fisher's Information, leads to $\pi(r) \propto 1/r^2$ (follows from Equation \ref{eq:jeffreys}), which has a similar tail behavior to the standard Half-Cauchy $\mathrm{C}^{+}$(0,1). Traditional Bayesian models often use an Inverse-Gamma prior for variance components. However, the Inverse-Gamma prior may induce strong shrinkage, especially when the data is limited \citep{gelman2006prior}. The Half-Cauchy prior, in contrast, is less informative and has been shown to perform better in many hierarchical settings, particularly when the number of groups is small. It avoids the problems associated with overly tight priors, which can lead to underestimating variability \citep{Polsonhalfcauchy}. \\

\begin{remark}
    With $r \sim \mathrm{C}^+(0,1)$ one can see that for $x>0$, \begin{align}\label{eq:RGcauchy}
\mathbb{P}_{\mathrm{HC}}(r>x\,|\,\omega)
&=\frac{1}{Z_{\mathrm{HC}}(\omega)}
  \int_{x}^{\infty}\frac{1}{1+r^{2}}
  \exp\!\left(-\frac{(\omega-1/r)^{2}}{2\sigma_{\omega}^{2}}\right)\,dr \nonumber\\
&\le \frac{1}{Z_{\mathrm{HC}}(\omega)}
  \int_{x}^{\infty}\frac{1}{r^{2}}
  \exp\!\left(-\frac{(\omega-1/r)^{2}}{2\sigma_{\omega}^{2}}\right)\,dr \nonumber\\
&=\frac{Z_{\mathrm{RG}^{+}}(\omega)}{Z_{\mathrm{HC}}(\omega)}\;
  \underbrace{\frac{1}{Z_{\mathrm{RG}^{+}}(\omega)} \int_{x}^{\infty}\frac{1}{r^{2}}
  \exp\!\left(-\frac{(\omega-1/r)^{2}}{2\sigma_{\omega}^{2}}\right)\,dr}_{=\;\mathbb{P}_{\mathrm{RG}^{+}}(r>x\,|\,\omega)} \nonumber\\
&=: c(\omega)\,\mathbb{P}_{\mathrm{RG}^{+}}(r>x\,|\,\omega),
\end{align}
where we define
\[
c(\omega):=\frac{Z_{\mathrm{RG}^{+}}(\omega)}{Z_{\mathrm{HC}}(\omega)}\ge 1,
\quad\text{since }\;\frac{1}{1+r^{2}}\le \frac{1}{r^{2}} \ \text{ for all }r>0,
\]
using the normalizing constants
$Z_{\mathrm{HC}}(\omega):=\int_{0}^{\infty}(1+r^{2})^{-1}
\exp\!\left(-(\omega-1/r)^{2}/{2\sigma_{\omega}^{2}}\right)\,dr$, and 
$Z_{\mathrm{RG}^{+}}(\omega):=\int_{0}^{\infty}{r^{-2}}
\exp\!\left(-(\omega-1/r)^{2}/{2\sigma_{\omega}^{2}}\right)\,dr.$

Thus, for small values of $\omega$, the tail behavior of the posterior under a Half-Cauchy prior increasingly resembles that of the posterior under a Reciprocal Gaussian prior.
\end{remark}

\item \textbf{Weibull Prior}: The Weibull prior with scale parameter 1 is heavy-tailed when $\alpha<1$, exhibiting sub-exponential behavior which is useful for modeling tail risks. \citet{weibullibrahim} provides an in-depth treatment of extreme-value distributions, including Weibull priors for modeling survival times and hazard functions. 
    
\item \textbf{Product Half-Cauchy Prior}: We propose the Product Half-Cauchy prior, typically from the standpoint of further slowing down the rate of tail decay for the Half-Cauchy prior. The product of two Half-Cauchy random variables has the following density: 
    $$\pi(r) = \frac{2}{\pi^2} \cdot\frac{2 \log r}{r^2 -1}, \ \ r>0$$
 The tail behavior is given by $\nicefrac{2 \log r}{r^2}$ as $r \rightarrow \infty$, where the presence of the 
$\log r$ term in the numerator actually causes the tail to be thicker the Half-Cauchy distribution. The product of two Half-Cauchy distributions (say, $\tau$ and $\lambda$ for instance) is often used as a prior for variance components in a hierarchical model. This formulation is conceptually similar to the Horseshoe prior \citep{carvalho2010horseshoe}, because it effectively shrinks small values of $r$ towards zero while allowing large values of $r$ to remain large due to the heavy tails of the Half-Cauchy distributions. However, it differs from the Horseshoe prior in the sense that there is no separate global and local shrinkage parameter in this single observation scenario. We note here that the prior distribution for the local shrinkage parameter in horseshoe+ prior \citep{bhadra2015horseshoe+} has a Product Half-Cauchy distribution. 
\end{enumerate}

\begin{figure}[h]
   \centering
   \includegraphics[width=0.8\linewidth]{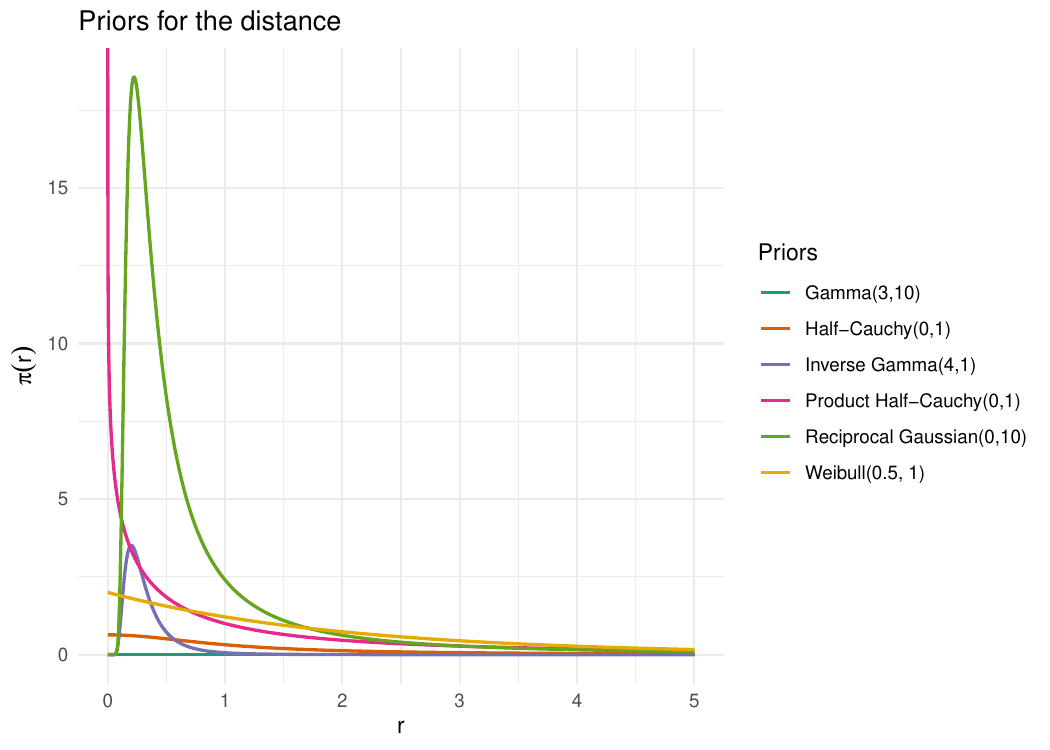}
   \caption{\footnotesize{Tail behavior of the candidate priors}}
    \label{fig:priorplots}
\end{figure}


These candidate priors can be sorted in terms of their tail-heaviness according to their p-credence criteria discussed in Lemma \ref{lem:1} (see Table \ref{tab:tail_decay}). Their tails can also be visualized through their density functions plotted in Figure \ref{fig:priorplots}. The lightest tail is exhibited by the Gamma$(3,10)$ prior, whereas the heaviest tail belongs to the Product Half-Cauchy$(0,1)$ prior.

\begin{table}[H]
    \centering
        \caption{Sorting the priors according to tail-heaviness (Light-Heavy) (\textit{top to bottom}).}
    \footnotesize{
    \label{tab:tail_decay}
    \begin{tabular}{lcl}
        \toprule
        \textbf{Prior} & \textbf{Tail Decay Rate} & \textbf{p-credence} \\
        \midrule
        Gamma$(n,\theta)$              & $e^{-\theta r}$ & $(1,\theta,n-1,0)$ \\
        Weibull$(c,1)$             & $e^{-r^c}$  & $(c,1,c-1,0)$ \\
        Inverse-Gamma$(\alpha,1)$      & $r^{-(\alpha + 1)}$ & $(0,0,-(\alpha+1),0)$ \\
        Reciprocal-Gaussian$(0,\sigma^2)$ & $r^{-2}$ & $(0,0,-2,0)$ \\
        Half-Cauchy$(0,1)$             & $r^{-2}$ & $(0,0,-2,0)$ \\
        Product Half-Cauchy$(0,1)$     & $\frac{\log(r^2)}{r^2}$ & $(0,0,-2,1)$ \\
        \bottomrule
    \end{tabular}}
\end{table}

\section{Theoretical Results}
\label{sec:theory}
\subsection{Tail Behavior of the Posterior}
In our endeavor to get robust Bayesian inference, limiting the influence of conflicting information (in our case, too small values of $\omega$), the use of heavy-tailed distributions is a valuable tool. It is natural to therefore characterize the tail behavior of the posterior through the notion of credence and p-credence in a bid to determine the dominance of the prior or the likelihood. The prior class which greatly simplifies our inverse-mean modeling problem is the class of \textit{reciprocal-invariant} priors. Reciprocal random variables $X$, where $X \ge 0$ by definition, satisfy the strong invariant property $X \stackrel{d}{=}X^{-1}$. They have been characterized by the following two equivalent conditions \citep{Hurlimann11}:
\begin{itemize}
    \item[(1)] The functional equation $$f(x) = \frac{1}{x^2}f\left(\frac{1}{x}\right), \ \ x>0.$$
    is satisfied for $f$ (density of $X$).
    \item[(2)] The transformed random variable $Y=\psi(X)$ is a symmetric random variable with respect to the origin, where $\psi:(0,\infty) \rightarrow \mathbb{R}$ is a monotone differentiable function with the property $\psi(x) = -\psi(x^{-1}),\ \ x>0$.
\end{itemize}
Note that both the Half-Cauchy and the Product Half-Cauchy priors fall under the category of reciprocal invariant. The simplification of the modeling problem is subject to the following observation that for Bayesian inference involving the distance parameter $r$, we look into the posterior $$\pi(r|\omega) \propto \pi(r)\exp\left(-\frac{1}{2 \sigma_{\omega} ^2} \left(\omega-\frac{1}{r} \right)^2 \right).$$
Under reciprocal invariance of $\pi(r)$, if we transform $r \mapsto t \coloneq 1/r$, the posterior density looks like
$$\pi(t|\omega) \propto \pi(t)\exp\left(-\frac{1}{2 \sigma_{\omega} ^2} \left(\omega-t \right)^2 \right).$$
which boils down to a Bayesian inference problem involving the location parameter $t$. Of course, this problem is well studied in the literature in the context of developing robust Bayesian procedures for guarding against sample outliers and prior mis-specifications. The following results help us understand the asymptotic behavior of the posterior density for the location parameter $t$ -- the term ``asymptotic" being used not in the frequentist sense, where the sample size goes to infinity, rather in the sense of having conflicting values, that is, $\omega \rightarrow 0$.

Throughout this section, we reparameterize by $t\coloneqq 1/r$ (can be interpreted as the \emph{true parallax}), so the likelihood is a Normal \emph{location} model in $t$, making it natural to impose and compare heavy–tailed priors via p-credence. Restricting ourselves to the class of reciprocal invariant priors, the following result enlightens us about the tail decay rate of the posterior, considering the influence of heavy-tailed priors, courtesy \cite{p-cred07}. 

\begin{theorem}\label{thm1}
    If $\text{p-cred}(\mathcal{L}(1/t;\omega)) = (\gamma',\delta',\alpha',\beta')$ and the $\text{p-cred}(\pi(t))=(0,\delta,\alpha,\beta)$, where we impose $\pi(.)$ to be reciprocal invariant, then $$\text{p-cred}(\pi(t|\omega)) = (\gamma',\delta',\alpha+\alpha',\beta+\beta')$$
    when $\gamma' > 0$.
\end{theorem}
Note that in our scenario, since we have a Normal likelihood, then $\pi \succ L$ for a sufficiently heavy-tailed prior $\pi(.)$ as $\gamma' > 0$ (in fact, $\gamma'=2$ to be exact). Theorem \ref{thm1} suggests that even under the influence of a heavy-tailed prior, the posterior tail behavior is dominated by the information source with higher credence or the first p-credence coefficient $\gamma'$ that is, the Normal likelihood. This dominance has a profound practical implication for our single-observation problem, which we formalize in the following corollary provided in \cite{o1990outliers}.
\begin{corollary}(Curse of a Single Observation)\label{corr1}
    \emph{The posterior probabilities for any fixed $t$ (or, $r$) are bounded by positive multiples of the likelihood. Specifically, for any given $d > 0$ and $\epsilon > 0$ such that $\forall \omega < A$ and $\forall t \le d$,
    $$\frac{k'}{K^{*}} (1-\epsilon) \le \pi(t|\omega) \left\{\mathcal{L}(1/t;\omega)\right\}^{-1} \le \frac{K'}{k^{*}}(1+\epsilon)$$
    with appropriate positive constants $k',K',k^{*},$ and $K^{*}$}.
\end{corollary}
In practical terms, this means that in situations where we have a single observation (even limited observations), the prior -- no matter how heavy-tailed, will not significantly alter the posterior distribution's behavior relative to the likelihood. The likelihood ``dominates" the posterior, rendering the prior's tail less impactful. We name this the \textit{``curse of a single observation"}!\\
Another manifestation of this curse is observed through the following proposition, which indicates that even if the prior decays slowly enough for large $r$, the posterior cannot put enough mass beyond ${1}/{\omega}$, or in fact, any function $c(\omega)$ which diverges as $\omega \rightarrow 0$.
\begin{proposition}\label{prop1}
   \emph{Let $\omega \mid r \sim \mathcal{N}(r^{-1},\sigma_{\omega} ^2)$ with fixed $\sigma_{\omega} \in (0,\infty).$ Under a reciprocally invariant prior $\pi_t$ on $t \coloneqq r^{-1}$, assuming that $\pi_t$ is locally integrable near $t=0$, then for any measurable function $c: \mathbb{R}^{+} \mapsto \mathbb{R}^{+}$ with $c(\omega) \rightarrow \infty$ as $\omega \downarrow 0,$
   \[
   \lim_{\omega \downarrow 0} \mathbb{P}(r > c(\omega) \mid \omega) = 0.
   \]
   Moreover, the rate depends on the local behavior of $\pi_t$ near $0$:}
\begin{enumerate}
       \item \textbf{Heavy-tailed case (finite credence).} \emph{If $\pi_t(t)\sim C_0\,t^\gamma$ as $t\downarrow 0$ for some $\gamma>-1$, then for $c(\omega)=A/\omega$ with $A>0$,
\[
\mathbb{P} \ \!\bigl(r>c(\omega)\mid \omega\bigr)\ =\ \frac{C_0\,\phi(0)}{(\gamma+1)A^{\gamma+1}Z(0)}\,\omega^{\gamma+1}\,[1+o(1)],
\]
where $\phi$ is the standard normal pdf and $Z(0)=\int_0^\infty \pi_t(u)\,\phi(u/\sigma_\omega)\,du\in(0,\infty)$.}
    \item \textbf{Light-tailed case (infinite credence).} \emph{Suppose the prior on $r$ has exponential-type tail, i.e., for some $R>0$, $c>0$, $\gamma>0$, $\alpha\in \mathbb{R}$ and $K<\infty$,
\[
\pi_r(r)\ \le\ K\, r^{\alpha}\,\exp(-c\,r^{\gamma}),\qquad r\ge R.
\]
Then, as $t\downarrow 0$,
\[
\pi_t(t)\ \le\ K\,t^{-(\alpha+2)}\exp\!\big(-c\,t^{-\gamma}\big),
\]
and for $c(\omega)=A/\omega$ one has the stretched–exponential decay
\[
\mathbb{P} \ \!\bigl(r>c(\omega)\mid \omega\bigr)\ \le\ C\,\omega^{\gamma-(\alpha+1)}\,\exp\!\big(-c'\,\omega^{-\gamma}\big)\,[1+o(1)],
\]
for some finite constants $C,c'>0$ (depending on $A,c,\gamma,\alpha,\sigma_{\omega}$). In particular, it vanishes faster than any power of $\omega$.}
   \end{enumerate}
\end{proposition}
The proofs of Theorem \ref{thm1} and Proposition \ref{prop1} are detailed in the Appendix \ref{sec:proofs}. 
The threshold $c(\omega)=A/\omega$ corresponds to distances larger than a fixed multiple of the naive inverse–parallax, i.e. $r>A/\omega$. Proposition \ref{prop1} says that, as the observed parallax $\omega\downarrow 0$, the posterior probability of such extreme distances \emph{must} vanish; no prior can keep substantial mass there once the likelihood says the parallax is small. What the prior controls is only the \emph{rate} of vanishing, through its local behavior near $t=0$ (equivalently, near infinite distance): polynomial behavior $\pi_t(t)\asymp t^\gamma$ yields a polynomial rate $\omega^{\gamma+1}$, whereas exponential-type priors force a much sharper, stretched–exponential decay. Thus heavy-tailed (polynomial) priors can \emph{delay} the collapse of posterior mass at very large $r$ (leaving more posterior allowance for large distances when $\omega$ is small but nonzero), but they cannot prevent that collapse as $\omega\to 0$. This is another face of the \emph{single-observation curse}: in the small-$\omega$ regime the Normal likelihood in $t$ ultimately dictates the posterior tail beyond any diverging threshold. 


\subsection{Posterior Risk and Prior Choices}\label{sec:risk}

We now broaden our scope beyond reciprocal invariant priors, exploring a wider class of non-degenerate priors and their influence on posterior risk under various loss functions. From a decision-theoretic perspective, it is critical to study the posterior risk, particularly under common loss functions such as squared error, absolute error, and others. Of special interest is the posterior squared-error risk: 
\[
R_{\beta}(r_0) = \int_{0}^{\infty} (r-r_0)^2 \pi_{\beta}(r) k(r) dr,
\]
where $k(r) = \exp(-A/r^2 +2A \omega/r)$, with $A=1/(2\sigma_{\omega} ^2)$, and $r_0$ is the true stellar distance. It has been seen in \citet{Bailer_Jones_2015}, that $R_{\beta}(r_0)$ diverges to infinity as $r_0 \rightarrow \infty$ regardless of the prior choice. However, the tail heaviness of the prior can significantly affect how rapidly this divergence occurs. Heavy-tailed priors accommodate large distances more naturally, thereby delaying this explosion.



In fact, we can relate the p-credence ordering of the priors and the corresponding posterior risk explosions. For every $\beta > -1$, we define \[
\pi_{\beta}(r) = \frac{(\log r)^{\beta}}{Z_{\beta} r^2} \mathbb{I}_{(1,\infty)}(r) + c_0 (r) \mathbb{I}_{(0,1]}(r)
\]
where $c_0 (r)$ is any bounded positive density on $(0,1]$ and $Z_{\beta} = \int_{1}^{\infty} r^{-2}(\log r)^{\beta} dr = \Gamma(\beta +1) < \infty$. Note that, all these priors share p-credences indexed by $\beta$, given by the tuple $(0,0,-2,\beta)$. We focus on priors with p-credence $(0,0,-2,\beta)$ because, per Proposition~\ref{prop1}, finite credence (polynomial) tails are the only ones that can \emph{delay} posterior tail collapse, and by fixing the polynomial order $r^{-2}$ while varying only the logarithmic refinement $\beta$, we obtain a clean one-parameter family to compare heavy tails on a common scale and isolate their effect on risk blow-up.

The following result precisely characterizes the influence of $\beta$ on the divergence rate of posterior risk:
\begin{theorem}\label{thm:pcredrisk}
\begin{itemize}
For every non-degenerate prior $\pi_{\beta}$, with $\beta > 0$ and all $r_0 \ge 2b > 2,$
\item[(i)] For constants $a,b$ chosen such that $1<a<\exp(\exp(-\gamma))<b$ with $\gamma \approx 0.58$ being the Euler--Mascheroni constant, and $c_k = \min_{r \in [a,b]} k(r) > 0$. 
\[
R_{\beta}(r_0) \ge C_{\beta}r_0 ^2, \quad C_{\beta}=\frac{c_k(a^{-1}-b^{-1})}{4 \Gamma(\beta+1)} (\log a)^{\beta}
\]
\item[(ii)] The map $\beta \mapsto C_{\beta}$ is strictly decreasing on $(0,\infty)$, i.e., for $\beta_1 < \beta_2 \implies C_{\beta_1}>C_{\beta_2}$.
\end{itemize}
\end{theorem}
This result suggests that if the p-credence of two prior densities share $(\gamma,\delta,\alpha)$ but have $\beta_1 < \beta_2$, then the prior with higher logarithmic decay $\beta_2$ will postpone the $r_0 ^2$ divergence the longest. Henceforth, we confirm our prior hierarchy through $\beta_{IG} < \beta_{HC} = \beta_{RG} < \beta_{PHC}$ (see Table \ref{tab:tail_decay}). 
\paragraph{Extension to Polynomially Dominated Loss Functions.} The above result naturally extends to polynomially dominated loss functions, which include squared error, absolute error, and Huber losses. Specifically, if we suppose that the loss function $\delta(r,r_0)$ satisfies the polynomial domination condition: $\exists \ p>0, d>0,$ such that $\delta(r,r_0) \ge dr_0 ^p$ for all $r \in [a,b]$, with $r_0 \ge 2b$, then a similar divergence result holds:
\[
R_{\beta,\delta}(r_0) \ge \tilde{C}_{\beta,\delta} r_0 ^p, \ \text{with} \ \tilde{C}_{\beta,\delta} \propto \frac{(\log a)^{\beta}}{\Gamma(\beta+1)}.
\]
Thus, heavier-tailed priors similarly delay the explosion of posterior risk under polynomially dominated loss functions.

\paragraph{Bounded loss functions.} For bounded loss functions, such as the fixed-width rejection loss defined as $\delta(r,r_0)=\mathbb{I}\{|r-r_0| > \epsilon\},$ we obtain the following result (Theorem \ref{thm:bounded_rejection_loss}). We illustrate the bounded-loss case with the fixed-width rejection loss because it is the canonical bounded binary criterion for distance accuracy, and its asymptotic risk rates are representative of \emph{any} bounded loss; extensions to other bounded losses follow by the same arguments.

\begin{theorem}\label{thm:bounded_rejection_loss}
Fix an observed parallax $\omega$ with error $\sigma_{\omega}$. As $r_0 \rightarrow \infty,$ we classify the posterior risk under the tail behavior of our priors as follows:
\begin{itemize}
    \item[(i)] \textbf{Exponential Tail:}  For exponential tail priors $\pi_{\text{exp}}(r) \propto r^{\alpha} \exp(-cr)$ with p-credence coefficient $\gamma > 0, c> 0$, and $\alpha>-1$, posterior risk converges exponentially fast to $1$:
    \[
    R_{\text{exp},\epsilon}(r_0) = 1-\mathcal{O}(r_0 ^{\alpha} \exp(-cr_0))
    \]
    \item[(ii)] \textbf{Polynomial Tail:} For polynomial tail priors with p-credence coefficient $\gamma=0$, and $p>1$, the posterior risk converges polynomially to 1:
    \[
    R_{\text{poly},\epsilon}(r_0) = 1-\Theta(r_0 ^{-p}(\log r_0)^{\beta})
    \]
\end{itemize}
\end{theorem}
Thus, for a fixed tolerance $\epsilon$, as the true distance $r_0$ gets larger, the posterior almost surely misses the $\epsilon-$ window around $r_0$, so the rejection risk tends to $1$. Exponential tails drive this collapse \emph{exponentially} fast, whereas heavier tails collapse only \emph{polynomially} fast, cushioning the deterioration.

\textbf{No prevention, only delay.} Taken together, Theorems~\ref{thm:pcredrisk} and \ref{thm:bounded_rejection_loss} show that posterior risk must diverge as $r_0\to\infty$; the prior only controls \emph{how fast}. Heavy-tailed (finite-credence) priors slow this growth: typically to polynomial rates, whereas, exponential-type tails accelerate it. This aligns with Proposition~\ref{prop1}: in the small-parallax regime the likelihood ultimately forces tail collapse, and heavy tails merely postpone it.

\paragraph{Regularity conditions underpinning robust tails.} To further clarify ``how heavy is sufficiently heavy," \citet{angers07} has quantified the thickness and regularity for the prior tails needed for robust location parameter inference, through three regularity conditions:

\begin{itemize}\label{regularity}
    \item [(C1)] For any $\epsilon, h > 0$, there exists a constant $A_1 (\epsilon,h) > 0$, such that $z > A_1(\epsilon,h)$ and $|\theta| \le h$ $\implies 1-\epsilon \le {\pi(z+\theta)}{\pi(z)}^{-1} \le 1+\epsilon$. 
    \item [(C2)] There exists a constant $M_2 > 1$ and a proper density $g$ such that for all $z > A_2$, 
    $$\frac{\pi^2 \left(\frac{z}{2}\right)}{\pi(z)g\left(\frac{z}{2}\right)} \le M_2.$$
    \item [(C3)] $$\frac{d^2}{dz^2} \log \pi^{*}(z) \ge \frac{d^2}{d z^2} \log g(z) \ge 0,$$
    where, $\pi^{*}$ can be the density $\pi$ or a density with the same tail behavior.
\end{itemize}
\begin{remark}
The condition (C1) implies that a location transformation has no impact on the right tail of the prior density $\pi(z)$ as $z \rightarrow \infty$. For instance, if $\pi(z)$ is the density of a Half-Cauchy distribution, $\lim_{z \rightarrow \infty} {\pi(z+\theta)}{\pi(z)}^{-1} = 1$. For conditions (C2) and (C3), we can take $g$ to be a double Pareto density as professed by \citet{p-cred07}.  
Condition (C3) ensures that the logarithm of the densities $\pi^{*}$, and $g$ have convex right tails, with the log-convexity of $\pi^{*}$ being more pronounced than that of $g$. It is worthwhile to note that among our list of candidate priors, these regularity conditions are satisfied by the Inverse Gamma, the Reciprocal Gaussian, the Half Cauchy, and the Product Half Cauchy priors.
\end{remark}

\section{Simulation Experiments}\label{sec:simuls}
We define a simulation setup for assessing the performance of our candidate priors discussed in Section \ref{sec:method} to estimate distances from the parallax measurements. We use $r_0$ to denote the true, data-generating distance throughout and examine how the prior choice affects distance estimation under two regimes:

\subsection{Setup A (fixed, small noise).}
To be specific, we compare the priors $\text{Gamma}(3,10)$, $\text{IG} (4,1)$, $\text{RG}^{+}(0,10)$, $\text{Weibull}(0.5,1)$, $\text{Half-Cauchy}(0,1)$, and $\text{Product Half-Cauchy}(0,1)$.  We generate a sequence of true parallaxes $\omega_{0j}$'s of length $J=500$, uniformly spaced between $0$ and $8$. The true values of the distances $r_{0j}$ is taken to be $1/\omega_{0j}$ for each instance. We fix $\sigma_{\omega}=0.045 \simeq {1}/{\sqrt{500}}$, based on the suggestions of \citet{Bailer_Jones_2015}. For practical purposes, it is generally advisable to take $\sigma_{\omega}$ depending primarily on the number of photons (N) received by the telescope from the star, or rather, its brightness. However, it has been observed that the choice of the noise parameter $\sigma_{\omega}$ plays a crucial role in determining the robustness and reliability of the model fit. By choosing a low value for $\sigma_{\omega}$, specifically $0.045$, our intent is to minimize the impact of observational noise on the model's performance -- allowing for a clearer evaluation of the priors' effectiveness in estimating distances without the confounding effects of large measurement errors. We also study the sensitivity of the model fit to different values of $\sigma_{\omega}$ recognizing that heavy-tailed priors react differently to varying noise levels (see Appendix \ref{sec:sensitivity}).

\begin{figure}[!ht]
  \centering
  \includegraphics[width=0.6\linewidth]{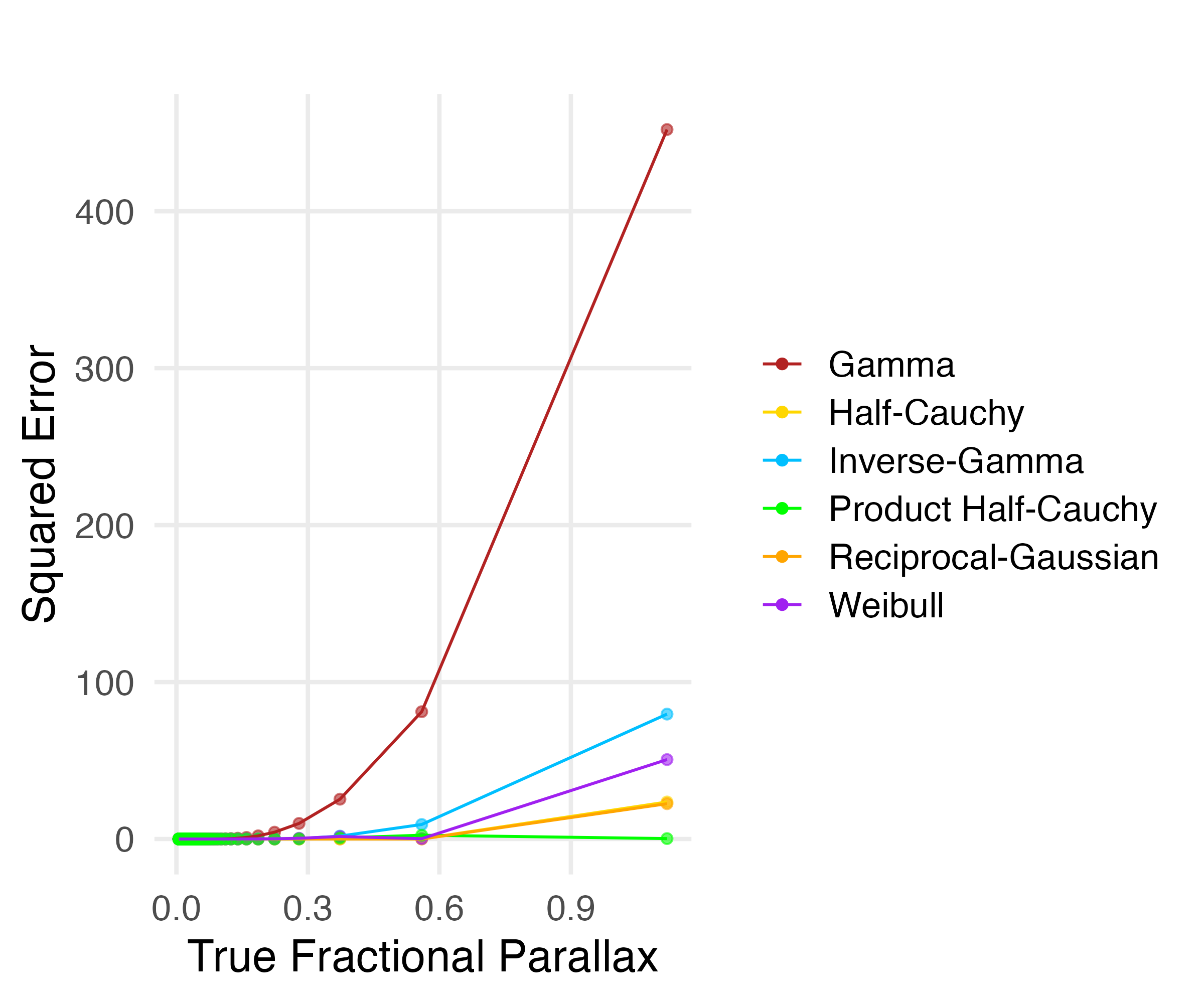}
  \caption{\footnotesize{Plot indicating how the squared error varies with the true distance across all the candidate priors, for $\sigma_{\omega}=0.045$.}}
  \label{fig:errorcomp}
\end{figure}

Then, we fit the hierarchical model: \[\omega_j \sim \mathcal{N}\left(\frac{1}{r_j},\sigma_{\omega} ^2\right), \quad r_j \sim \pi(r_j).\]
Our goal in this experiment is to measure the squared error $(\hat{r}_j-r_{0j})^2$ as a function of the true value of the fractional parallax error ($f$), which is defined as $\sigma_{\omega}.r_{0j}$, a measure of the relative error in the parallax estimate (see Figure \ref{fig:errorcomp}).  Note that we take the sequence of observed parallaxes $(\omega)$'s in such a way that $f$ can vary widely from $0.005$ to $1$. Of course, we are interested in studying the behavior of the discrepancy between $\hat{r}_j$ and $r_{0j}$, when the fractional parallax error increases (or equivalently, large values of $r_{0j}$).  
As seen in Figure \ref{fig:errorcomp}, the Product Half-Cauchy prior exhibits the smallest growth in squared error—essentially flat as the true distance increases--consistent with its very heavy (log-polynomial) tails. The Half-Cauchy and Reciprocal-Gaussian behave similarly, with only modest increases in error, matching their shared $r^{-2}$ tail behavior discussed in Section \ref{sec:method}. In contrast, the Gamma prior deteriorates sharply once $f\gtrsim 0.4$, reproducing the instability previously reported for exponentially decaying distance priors \citep{Bailer_Jones_2015}.

It is important to note that the robustness of the Product Half-Cauchy, Half-Cauchy and the Reciprocal-Gaussian priors extends to large fractional parallax error thresholds (the parallax error being larger than $0.8$). However, such extreme error thresholds are generally not encountered in practice due to the high precision of modern instruments. This robustness is more of a theoretical advantage, ensuring stability and reliability of the estimation process even under conditions of unusually high measurement errors.
\subsection{Setup B (varying noise).}
We simulate true distances from a constant volume density following the idea from \citet{Bailer_Jones_2015}:
\[
r_{0j} \sim \pi_{CV}(r) \propto r^2\,\mathbb{I}_{(0,10^4]}(r),\ j=1,\dots,500.
\]
For a grid of large fractional error levels $f\in[0.6,2.0]$, we draw
\[
\omega_j \sim \mathcal{N}\Big(\tfrac{1}{r_{0j}},\,(\tfrac{f}{r_{0j}})^2\Big),
\]
so that the parallax signal-to-noise ratio is controlled by $f$. We fit the same Bayesian models and compute two posterior risk criteria aggregated over $j$:
\[
\mathrm{RMSE}(f) \equiv \Big(\tfrac{1}{500}\sum_{j}(\hat r_j-r_{0j})^2\Big)^{1/2}, 
\
\mathrm{MSRE}(f) \equiv \Big(\tfrac{1}{500}\sum_{j}(1-\hat r_j/r_{0j})^2\Big)^{1/2},
\]
i.e., a distance-scale error (RMSE) and a scale-free mean-squared \emph{relative} error (MSRE). Note that, we fit Bayesian models using the following six candidate priors, each with a scale of $1000$ pc to reflect typical galactic distances: Gamma$(3,1000)$, IG$(3,1000)$, Half-Cauchy$(0,1000)$, RG$^{+}(0,1000)$, Weibull$(0.8,1000)$, and Product Half-Cauchy$(0,1000)$.
 \begin{figure}
    \centering
    \subfloat[$\delta(\hat r,r_0)=(\hat r-r_0)^2$]{\includegraphics[width=0.5\linewidth]{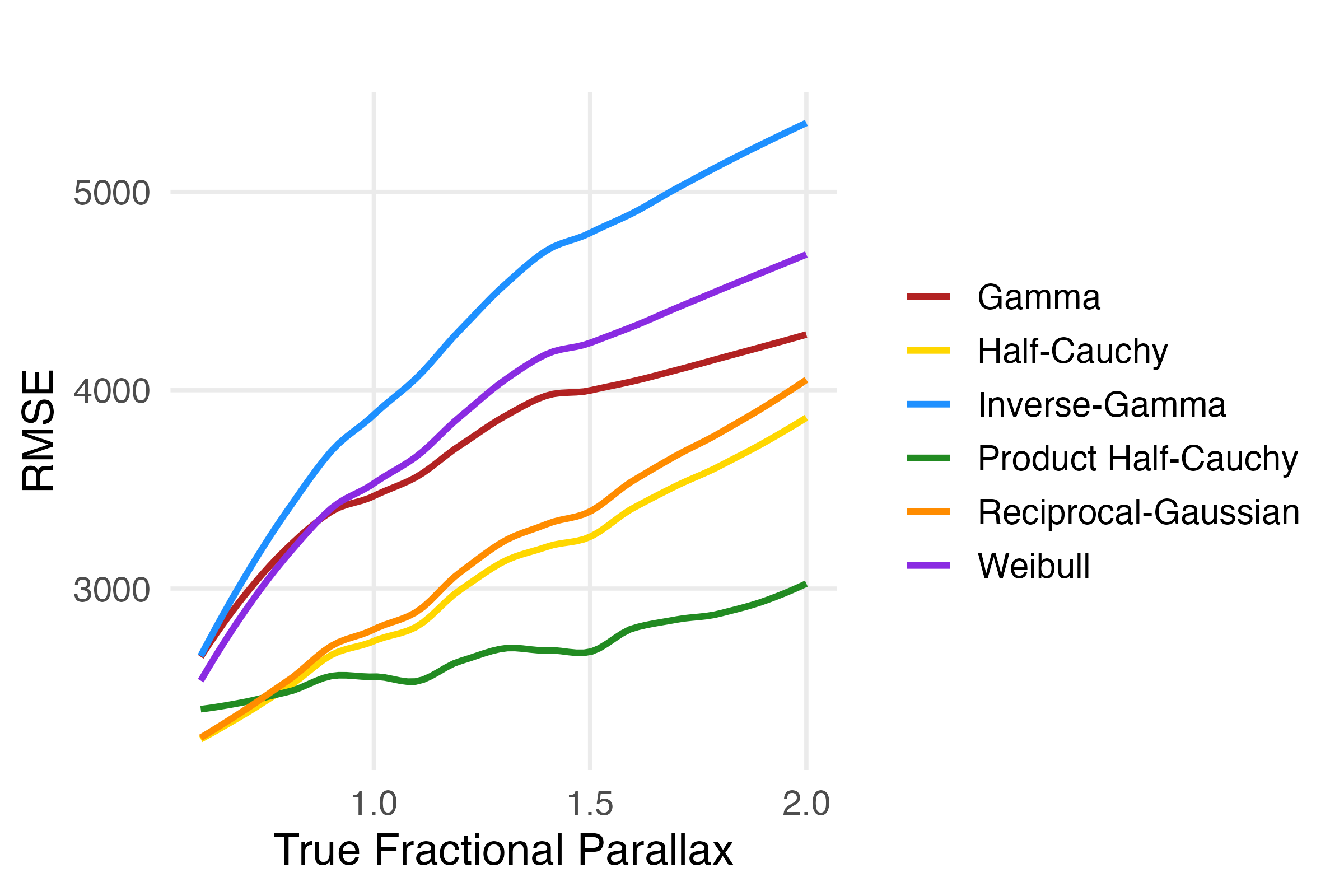}}
    \hfill
    \subfloat[$\delta(\hat r,r_0)=(1-\hat r/r_0)^2$]{\includegraphics[width=0.5\linewidth]{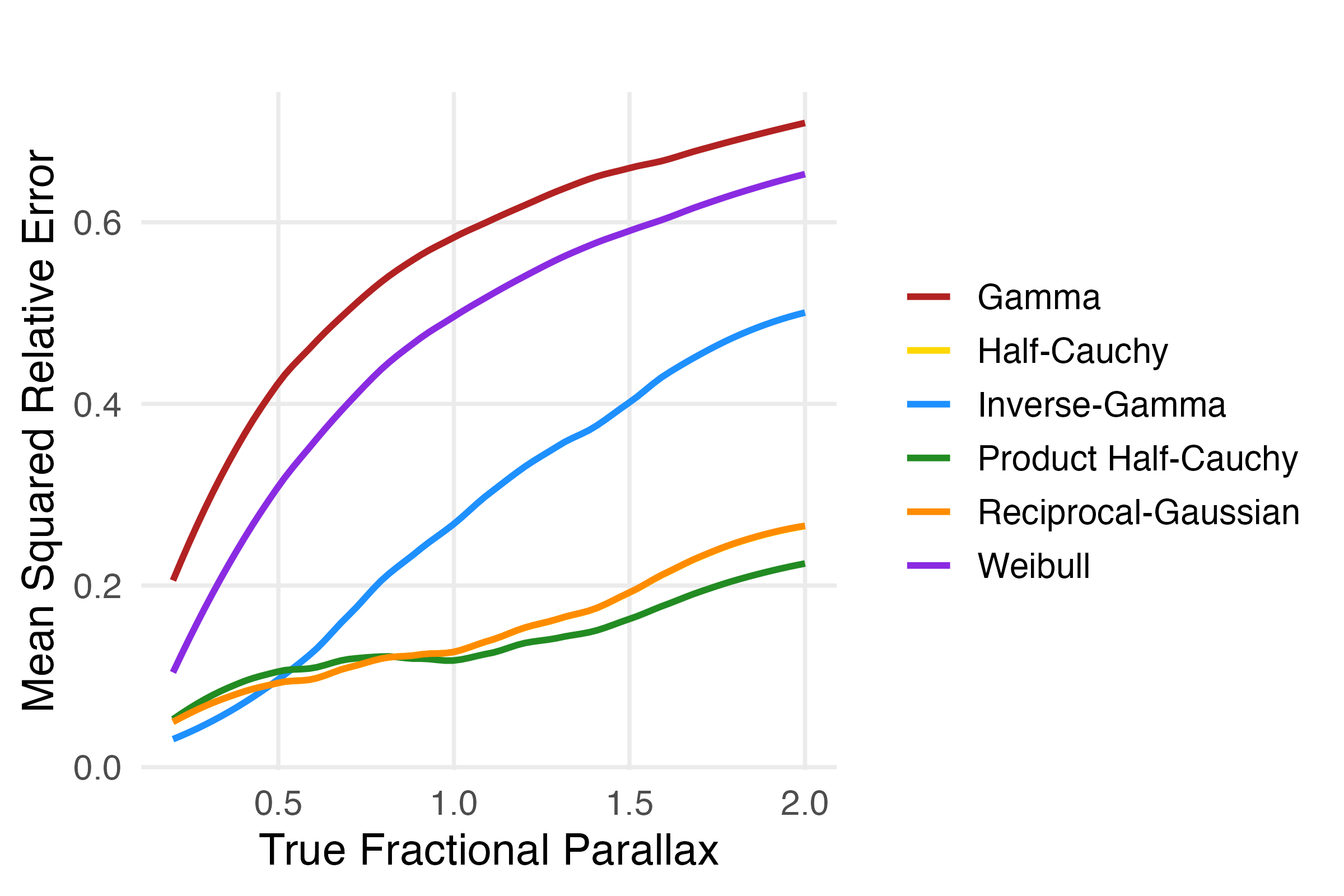}}
    \caption{\footnotesize{Posterior risk explosion under two different loss functions with varying $\sigma_{\omega}$.}}
    \label{fig:simchallenging}
  \end{figure}
Figure \ref{fig:simchallenging} displays the results from this more challenging simulation setting, where the fractional parallax error is high, clearly distinguishing the performance of the priors. The left panel shows that the Product Half-Cauchy prior achieves the slowest Root Mean Squared Error (RMSE) blowup across high fractional parallax error levels, demonstrating its superior ability to control the absolute error of the distance estimates. In the right panel, which considers the scale-free Mean Squared Relative Error (MSRE), both the Product Half-Cauchy and Reciprocal-Gaussian priors perform better than the rest, indicating they are most effective at producing proportionally accurate estimates. Their similar performance under the highly sensitive relative error metric is likely due to the similarity of their underlying tail structures, as discussed in Table \ref{tab:tail_decay}. In both scenarios, the heavy-tailed priors consistently outperform the lighter-tailed Gamma and Weibull priors. 

\paragraph{Implementation details.}
We compute posteriors using Hamiltonian Monte Carlo and no U-turn sampling (HMC - NUTS) \citep{hoffman2011nouturnsampleradaptivelysetting} via \texttt{R-stan} \citep{carpenter2017stan}. Each fit uses 5{,}000 iterations with 2{,}000 warmup for a single chain, and we summarize by the posterior median $\hat r$ (posterior means can be unstable or undefined under very heavy tails, e.g., Reciprocal–Gaussian, Half–Cauchy, Product Half–Cauchy). In this one–parameter setting the posterior normalizing constant can indeed be obtained by accurate one–dimensional numerical integration, and either approach should yield essentially identical point summaries such as posterior medians. We nevertheless use MCMC for two main reasons: (i) it produces full posterior draws that we use for uncertainty quantification (credible intervals, posterior predictive checks, loss summaries), even though we do not dwell on these here; and (ii) it scales seamlessly to natural extensions considered elsewhere (unknown $\sigma_\omega$, hierarchical models, mixture or hyperpriors), where low–dimensional quadrature ceases to be practical. Importantly, none of our qualitative conclusions depend on using MCMC rather than numerical integration in the present experiments. The \textsc{R} and \textsc{Stan} codes for reproducing the results in this paper are available at \href{https://github.com/DattaHub/inversemean/}{https://github.com/DattaHub/inversemean}. 

\section{Real Data Results}\label{sec:realdata}
We utilize an elaborate astrometric dataset extracted from the \textit{Gaia} Data Release 1 (GDR1), which encompasses positional, parallactic, and proper motion measurements for approximately 2 million of the brightest stars. These stars are included in both the \textit{Hipparcos} and \textit{Tycho-2} catalogues, providing a rich source of data for precise astronomical studies. For GDR1, a complex mathematical model known as the Global Astrometric Solution (\cite{GAS-GDR1}) was used to combine the \textit{Gaia} observations with the earlier Hipparcos and Tycho-2 data. This method ensures consistency across different datasets and helps in minimizing systematic errors. Moreover, \textit{Gaia}’s observations over multiple years (and \textit{Hipparcos} before it) provide a long baseline for measuring parallax shifts. A longer time baseline enhances the precision of parallax measurements, as the small apparent movements of stars due to the Earth’s orbit around the Sun can be more accurately detected.

Our analysis focuses on computing the posterior distances for a selected sample of 100 stars. By comparing the medians of the posterior distributions derived from these priors to the naive inverse parallax estimates provided in the GDR1 database, we aim to assess the accuracy and reliability of the proposed hierarchical models. It is to be noted that the parallaxes are in mas (milli-arcseconds), and angles are in degrees in the original database. We convert the parallaxes to arcseconds so that distances come out as pc (parsecs). To make the scale of our priors consistent with the magnitude of the data and for fair comparison across the priors, we choose the common scale parameter $L=1000$.


\begin{figure}[H]
    \centering
    \begin{subfigure}[b]{0.48\linewidth}
        \centering
        \includegraphics[width=\linewidth]{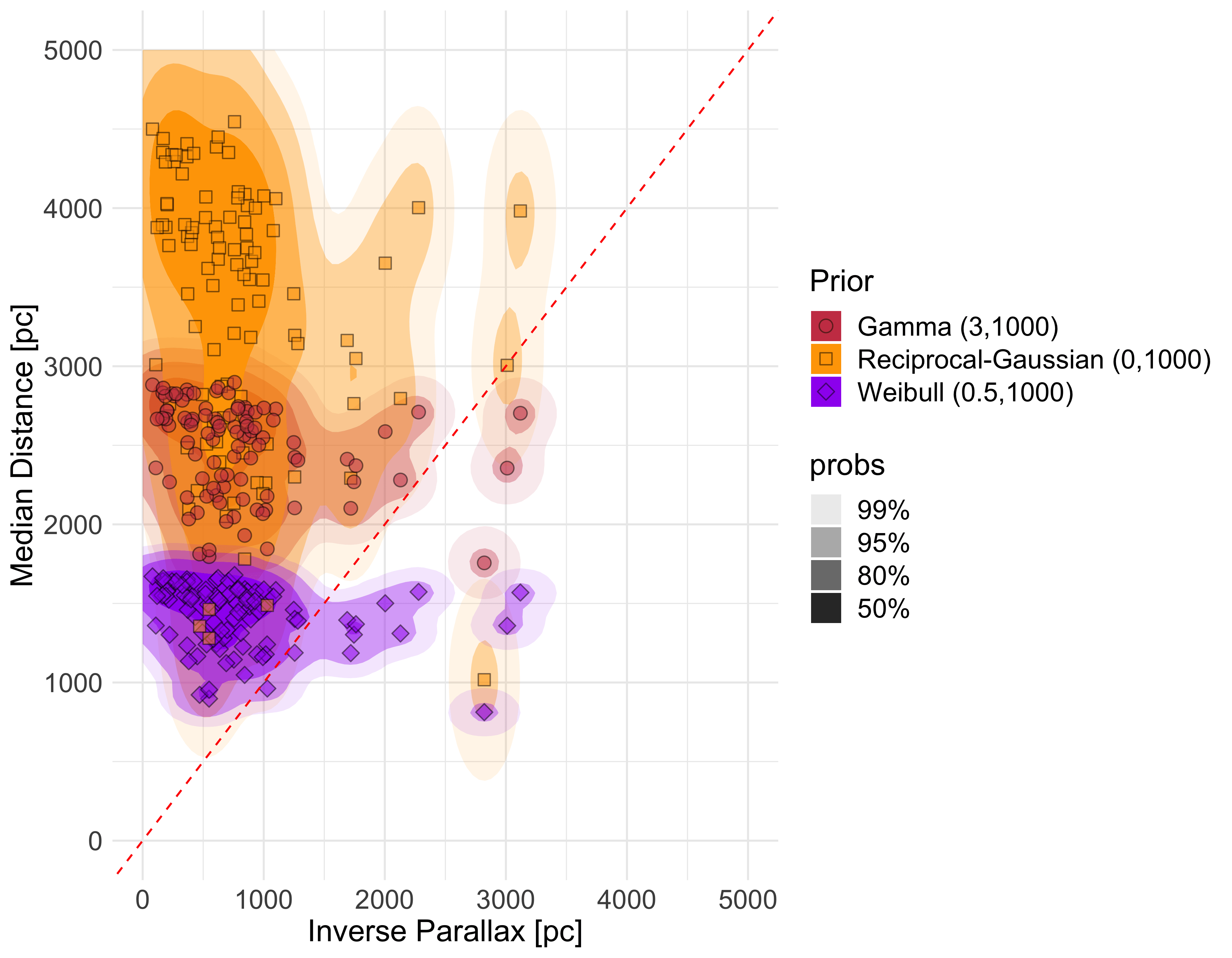}
        \caption{Exponential-type decay priors}
        \label{fig:light_tailed}
    \end{subfigure}
    \hfill
    \begin{subfigure}[b]{0.5\linewidth}
        \centering
        \includegraphics[width=\linewidth]{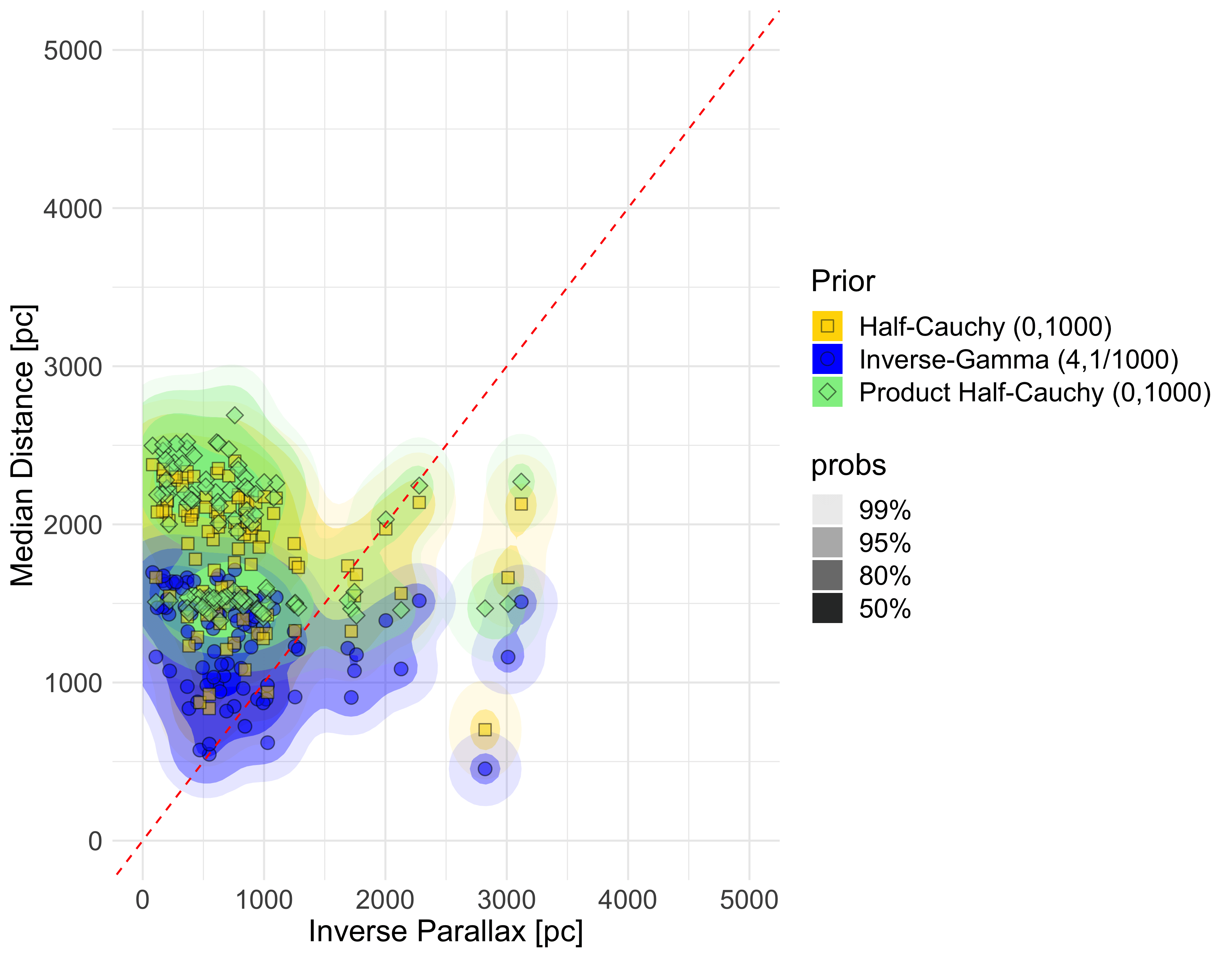}
        \caption{Polynomial-type decay priors}
        \label{fig:heavy_tailed}
    \end{subfigure}
    \caption{\footnotesize{Plotting the posterior estimates against the naive GDR1 Inverse Parallax estimates.}}
    \label{fig:4}
\end{figure}
The plot in Figure \ref{fig:light_tailed} contrasts the posterior median estimates against the tabulated GDR1 naive inverse parallax estimates for priors whose tail decay rates are exponential (see Table \ref{tab:tail_decay}). Similarly, Figure \ref{fig:heavy_tailed} enlists and compares all the priors with heavier tails, their tail decay rates being polynomial. The degree to which points cluster around the diagonal line (red dotted line in each figure) indicates how closely the Bayesian estimates align with the inverse parallax estimates. A tight clustering around the diagonal suggests strong agreement, while a broad scatter indicates discrepancies between the two methods. The Gamma and Reciprocal-Gaussian priors have their densest regions overlapping the naive estimates more closely, while the Weibull prior shows a broader spread with a significant chunk of  its 50\% confidence region lying below the identity line. This indicates that the Weibull prior might be more skeptical of the larger distance estimates provided by the naive method. The Reciprocal-Gaussian prior has the broadest spread amongst the three light-tailed priors, suggesting that it might be allowing for more variability in the posterior estimates, especially at higher distances.  

The Half-Cauchy prior tends to shrink the distance estimates compared to the naive method, especially for stars with larger parallax (shorter distances), whereas the contours for the Inverse-Gamma prior suggest that it allows for a broader range of possible distances, especially at lower values. The Product Half-Cauchy prior appears to be more conservative, keeping estimates closer to the identity line but still showing a tendency to shrink distances compared to the naive method. 


All the candidate priors produce posterior median estimates that do not perfectly align with the naive inverse parallax estimates. This misalignment is expected, as the Bayesian models incorporate additional prior information and are not solely dependent on the parallax data -- which naturally introduce biases, especially in cases of large parallax errors or when the parallax is small.
Polynomial-tailed priors are usually more appropriate when dealing with such scenarios, offering a more cautious approach, which can be beneficial in ensuring that the posterior estimates are not unduly influenced by extreme values.

\section{Discussion}
\label{sec:disc}

In this study, we explored the challenge of estimating stellar distances from parallax measurements, focusing on how prior choice shapes Bayesian inference for the Normal inverse–mean problem. The reciprocal (nonlinear) map $r=1/\omega$ induces highly skewed, long–tailed behavior for distance when $\omega$ is small, so robustness is essential. We brought together the notions of credence and p–credence to compare tail thicknesses and to formalize when the likelihood or the prior dominates the posterior. Working with reciprocal–invariant priors lets us reparameterize to $t=1/r$ and reduce the problem to a well–studied robust location model.

Two results drive our conclusions. First, Theorem \ref{thm1} shows that with a Normal likelihood (p–credence first coefficient $\gamma'>0$), the posterior’s first p–credence coefficient coincides with that of the likelihood, irrespective of how heavy the prior tail is. This is our formal ``curse of a single observation": with one noisy parallax, the likelihood in $t$ dominates global posterior tail shape. Second, Proposition~\ref{prop1} quantifies posterior tail probabilities relevant for large distances: for any threshold $c(\omega)\uparrow\infty$ as $\omega\downarrow 0$, the posterior probability collapses to $0$. Moreover, the rate depends on local mass of the prior near $t=0$ (equivalently, large $r$): finite–credence (polynomial) priors yield polynomial decay in $\omega$, while infinite–credence (exponential–type) priors yield much faster, stretched–exponential decay. The risk results (Theorems~\ref{thm:pcredrisk} and \ref{thm:bounded_rejection_loss}) echo this message: heavy tails cannot \emph{prevent} risk blow–up as $r_0\to\infty$, but they can \emph{delay} it---slowing the approach to $1$ under bounded losses and reducing the leading constant under squared–error loss. 

Both simulation regimes support the theory. In Setup~A (fixed small noise), the Product Half–Cauchy (PHC)  maintains an almost flat error profile as $f=\sigma_\omega r$ grows, while Half–Cauchy  and Reciprocal–Gaussian track each other with modest growth; the Gamma prior deteriorates rapidly once $f\gtrsim 0.4$ (Figure~\ref{fig:errorcomp}). In the more challenging Setup~B with varying noise (large $f\in[0.6,2]$), PHC achieves the slowest RMSE blow-up, and similarly minimizes the relative error (MSRE) across the range (Figure~\ref{fig:simchallenging}). These ordering patterns align with Table~\ref{tab:tail_decay}: heavier polynomial (or log–polynomial) tails buy robustness at large $r$, whereas lighter exponential tails give up early.

On the \textit{Gaia} GDR1 subset, polynomial–tail priors produce conservative, shrinkage–aware estimates that remain close to the identity line while avoiding the most extreme inversions (Figure~\ref{fig:4}). The Half–Cauchy shrinks aggressively for nearby stars; the Inverse–Gamma allows broader lower–distance variability; and the Product Half-Cauchy balances both, staying near the naive $1/\omega$ while tempering extremes. 

We acknowledge that the ``curse’’ is about \emph{one} noisy parallax. With $n$ independent parallaxes, the likelihood in $t$ remains Normal but concentrates at rate $\sigma_\omega/\sqrt{n}$, so posterior mass outside any fixed neighborhood of the MLE (or the sample mean, $\bar\omega$) vanishes rapidly. In that regime the prior tail matters less for point summaries, and robustness questions shift from prior tails to modeling of outliers and to hierarchical pooling across many stars. The heavy–tail advantage does not disappear, but it becomes a second–order design choice relative to data quality and model mis-specification.

Broadening our future scope for investigation, the class of reciprocal–invariant priors satisfying (C1)–(C3) is not exhaustive. Compositions with similar tail shape—e.g., a product of Half–Cauchy and Reciprocal–Gaussian (a ``Reciprocal–Horseshoe’’)—inherit much of PHC’s robustness without strict invariance. Identifying broader classes of such composites is a promising direction. A second avenue is the choice of loss: squared error is natural for scale comparability but may be misaligned with scientific targets; our results under relative error and bounded losses already show different emphases. Exploring asymmetric or decision–relevant losses (e.g., luminosity–induced) (as seen in \citet{Bailer_Jones_2021}), which penalize errors in derived physical properties like stellar brightness more heavily, is a fruitful path.

In conclusion, heavy–tailed priors with polynomial or log–polynomial decay provide the most reliable behavior for parallax–based distance inference in the small–$\omega$ regime. They cannot avert the fundamental limitations of the inverse–mean geometry with a single observation, but they do \emph{delay} risk explosion and stabilize estimation where it matters most. Coupled with sensible scaling and hierarchical extensions, these priors form a robust and practical toolbox for precise mapping of the cosmos.

\appendix
\renewcommand{\thesubsection}{\thesection.\arabic{subsection}}

\section{Appendix}
\label{sec:appendix}
\subsection{Theoretical Proofs}\label{sec:proofs}
We present the proofs of Theorem \ref{thm1} and Proposition \ref{prop1} in this section. \\

\begin{proof}(Proof of Theorem \ref{thm1}) Without loss of generality assume that $z_0 > 0$ in (\ref{eq:gep}). The key observation here is the following, courtesy Lemma 1 in \cite{p-cred07}, \\
\emph{If $c \neq 0$, $d \in \mathbb{R}$, then $\forall z \in \mathbb{R}$, there exists constants $0<a \le A < \infty$ such that \begin{equation}\label{eq:lemma1}
    a \le \frac{\max(|z|,z_0)}{\max(|cz+d|,z_0)} \le A
\end{equation}}

Recall that the $\text{p-cred}(\mathcal{L}(1/t;\omega))=(\gamma',\delta',\alpha',\beta')$ and the $\text{p-cred}(\pi(t))=(0,\delta,\alpha,\beta)$. Define $m(\omega)$ to be the marginal distribution of $\omega$. Using the above result (\ref{eq:lemma1}) and the definition of p-credence, it's straightforward to observe that for constants $A,K,K' > 0$: 
\begin{align}\label{eq:upperbd}
    \pi(t|\omega) & = \frac{\mathcal{L}(1/t;\omega)\pi(t)}{m(\omega)} \nonumber\\
    & \le \frac{K K'}{m(\omega)} \max(|t|,z_0)^\alpha \max(|t-\omega|,z_0)^{\alpha'} \log^{\beta}[\max(|t|,z_0)] \nonumber\\
    & \log^{\beta'}[\max(|t-\omega,z_0)]\exp\left\{-[\delta' \max(|t-\omega|,z_0)^{\gamma'}]\right\} \nonumber\\
    & \le \frac{AKK'}{m(\omega)}\max(|t-\omega|,z_0)^{\alpha+\alpha'} \log^{\beta+\beta'}[\max(|t-\omega|,z_0)] \nonumber\\
    & \exp\left[-\delta' \max(|t-\omega|,z_0)^{\gamma'}\right] 
\end{align}
Similarly, we can have a similar lower bound using (\ref{eq:lemma1}), where there exists constants $a,k,k'>0$ such that: 
\begin{align}\label{eq:lowerbd}
    \pi(t|\omega) & \ge \frac{akk'}{m(\omega)} \max(|t-\omega|,z_0)^{\alpha+\alpha'} \log^{\beta+\beta'}[\max(|t-\omega|,z_0)] \nonumber\\
    & \exp[-\left\{\delta' \max(|t-\omega|,z_0)^{\gamma'}\right\}] 
\end{align}
Combining the bounds (\ref{eq:lowerbd}) and (\ref{eq:upperbd}), we have that $$\pi(t|\omega) \approx p(t-\omega|\gamma',\delta',\alpha+\alpha',\beta+\beta')$$
and the result follows.    
\end{proof}

\begin{proof}(\textbf{Proof of Proposition \ref{prop1}}) We write the posterior for $t$:
\[
p(t\mid \omega)=\frac{\pi_t(t)\,\phi\big((t-\omega)/\sigma_\omega\big)}{Z(\omega)},\qquad
Z(\omega)=\int_0^\infty \pi_t(u)\,\phi\big((u-\omega)/\sigma_\omega\big)\,du.
\]
For any threshold $c(\omega)$ let $\varepsilon(\omega)=1/c(\omega)\downarrow 0$. Then
\[
\mathbb{P}(r>c(\omega)\mid \omega)=\mathbb{P}(t<\varepsilon(\omega)\mid\omega)
=\frac{N(\omega)}{Z(\omega)},\ 
N(\omega)=\int_0^{\varepsilon(\omega)} \pi_t(t)\,\phi\big((t-\omega)/\sigma_\omega\big)\,dt.
\]

Analyzing the denominator, since $\pi_t$ is locally integrable near $0$ and $\phi$ is bounded, dominated convergence gives $Z(\omega)\to Z(0)=\int_0^\infty \pi_t(u)\,\phi(u/\sigma_\omega)\,du\in(0,\infty)$ as $\omega\downarrow 0$.

\emph{Finite credence case (i).} If $\pi_t(t)\sim C_0 t^\gamma$ as $t\downarrow 0$ with $\gamma>-1$, then on $t\in[0,\varepsilon(\omega)]$ we have $\phi((t-\omega)/\sigma_\omega)=\phi(0)[1+o(1)]$ and
\[
N(\omega)=\phi(0)\int_0^{\varepsilon(\omega)}\pi_t(t)\,dt\,[1+o(1)]
=\phi(0)\,\frac{C_0}{\gamma+1}\,\varepsilon(\omega)^{\gamma+1}[1+o(1)].
\]
With $\varepsilon(\omega)=\omega/A$ this yields the stated expansion; dividing by $Z(\omega)\to Z(0)$ gives (i). For a general $c(\omega)\to\infty$, $\varepsilon(\omega)\downarrow 0$ implies $N(\omega)=\mathcal O(\varepsilon(\omega)^{\gamma+1})$ and hence $N(\omega)/Z(\omega)\to 0$.

\emph{Infinite credence case (ii).} From the tail bound on $\pi_r$, for $t$ small
\[
\pi_t(t)=\frac{\pi_r(1/t)}{t^2}\ \le\ K\,t^{-(\alpha+2)}\exp\!\big(-c\,t^{-\gamma}\big).
\]
Thus,
\[
N(\omega)\le \phi(0)\int_0^{\varepsilon(\omega)} K\,t^{-(\alpha+2)}\exp(-c\,t^{-\gamma})\,dt.
\]
With the substitution $u=c\,t^{-\gamma}$ one obtains
\[
\int_0^{\varepsilon} t^{-(\alpha+2)}e^{-c t^{-\gamma}}\,dt
= \frac{1}{\gamma}\,c^{-(\alpha+1)/\gamma}\,\Gamma\!\Big(\frac{\alpha+1}{\gamma},\,c\,\varepsilon^{-\gamma}\Big),
\]
an upper incomplete gamma. As $\varepsilon\downarrow 0$, the lower limit blows up, and the standard asymptotic
$\Gamma(s,x)\sim x^{s-1}e^{-x}$ yields
\[
N(\omega)\ \le\ C_1\,\varepsilon(\omega)^{\,\gamma-(\alpha+1)}\,\exp\!\big(-c\,\varepsilon(\omega)^{-\gamma}\big)\,[1+o(1)].
\]
For $c(\omega)=A/\omega$ so that $\varepsilon(\omega)=\omega/A$,
\[
N(\omega)\ \le\ C_2\,\omega^{\,\gamma-(\alpha+1)}\,\exp\!\big(-c'\,\omega^{-\gamma}\big)\,[1+o(1)].
\]
Divide by $Z(\omega)\to Z(0)>0$ to obtain (ii). For any $c(\omega)\to\infty$ the same bound with $\varepsilon(\omega)\downarrow 0$ implies $N(\omega)/Z(\omega)\to 0$.

\end{proof}

\begin{proof}(\textbf{Proof of Theorem \ref{thm:pcredrisk}})
    \begin{itemize}
        \item[(i)] For the chosen constants $a,b$, since $r \in [a,b] \subset (0,r_0),$ we have
        \[
        (r-r_0)^2 \ge r_0 ^2 (1-b/r_0)^2 \ge r_0 ^2 /4 \quad  \forall r_0 \ge 2b
        \]
        We get \begin{align*}
            R_{\beta}(r_0) & \ge \int_{a}^{b} (r-r_0)^2 \pi_{\beta}(r)k(r) dr \\
            & \ge \frac{r_0 ^2 c_k}{4 \Gamma(\beta+1)} \int_{a}^{b}(\log r)^{\beta}r^{-2} dr \\
            & \ge \frac{r_0 ^2 c_k}{4 \Gamma(\beta+1)} (\log a)^{\beta} (a^{-1}-b^{-1}) \\
            & = C_{\beta} r_0 ^2
        \end{align*}
        \item[(ii)] Let $F(\beta) = \frac{(\log a)^{\beta}}{\Gamma(\beta+1)}$. Studying the monotonicity of $C_{\beta}= \frac{c_k(a^{-1}-b^{-1})}{4}F(\beta)$ is equivalent to studying $F$. Note that \[
        \log F(\beta) = \beta \log \log a - \log \Gamma(\beta+1)
        \]
        Differentiating w.r.t. $\beta$, we have \[
        \frac{d}{d \beta} \log F(\beta) = \log \log a - \psi(\beta+1)
        \]
        with $\psi(.)$ being the di-gamma function. Since $\psi(\beta)$ is increasing in $\beta$, $\psi(\beta+1) \ge \psi(1) = -\gamma, \quad \forall \beta>0$. For our choice of $a$, we indeed have $\log \log a + \gamma < 0$ and hence the map $\beta \mapsto C_{\beta}$ is decreasing.
    \end{itemize}
\end{proof}
\begin{proof}(\textbf{Proof of Theorem \ref{thm:bounded_rejection_loss}})
Throughout set the interval $I_0 \coloneqq [r_0 - \epsilon,r_0 + \epsilon]$. Also, define the two integrals $A_1 \coloneqq \int_{I_0} \pi(r)k(r) dr,$ and $A_2 \coloneqq \int_{I_0 ^{c}} \pi(r) k(r) dr.$ Then note that the posterior risk in either case can be written as $R_{\pi,\epsilon}(r_0)=1-A_1/(A_1+A_2)$. We now treat the two prior classes separately.
\paragraph{Exponential tail.} We upper bound $A_1$ first. For $r \in I_0,$ we have 
\[
\pi_{\text{exp}}(r) \le C e^{c\epsilon}(r_0 + \epsilon)^{\alpha}e^{-cr_0}
\]
Also, since $k(r) \le 1$, we have the upper bound for $A_1$ as 
\[
A_1 \le C e^{c\epsilon}(2\epsilon)(r_0 + \epsilon)^{\alpha}e^{-cr_0} = \mathcal{O}(r_0^{\alpha} e^{-cr_0})
\]
All it remains to show is a strictly positive lower bound for the denominator $A_1 + A_2.$ To see that, we choose any fixed $M>0$.Now,
\begin{align*}
    A_1 + A_2 & = \int_{0}^{\infty} k(r) \pi(r) dr \\
    & \ge \int_{M}^{2M} k(r) \pi(r) dr \\
    & \ge \exp \left(-A(M^{-1} + |\omega|)^2 \right) \int_{M} ^{2M} \pi(r) dr  \\
    & > 0
\end{align*}
The last inequality follows by observing that for every $r \in [M,2M],$ we have $|r^{-1} - \omega| \le M^{-1} + |\omega|$.
Thus, indeed we have $R_{\text{exp},\epsilon}(r_0) = 1-\mathcal{O}(r_0 ^{\alpha} e^{-cr_0}).$
\paragraph{Polynomial tail.} We first pick a fixed $\eta \in (0,1)$ and split $(0,\infty) = (0, \eta r_0) \cup [\eta r_0, \infty).$ Then, it is worthwhile to note that $k(r) = 1+ \mathcal{O}(r_0 ^{-2})$ on $I_0$.  This can be proved by letting $r=r_{0}x,$ with $x \in [\eta,\infty)$. Tracking back, we call $\omega_0 = r_0 ^{-1}$ as the true parallax. Since, $\omega \sim \mathcal{N}(\omega_0,\sigma_{\omega}^2),$ we can deduce $\omega = \mathcal{O}(\omega_0)=\mathcal{O}(r_0 ^{-1})$. Finally, $|r^{-1}-\omega| = |(r_0 x)^{-1} - \omega| \le (r_0 \eta)^{-1}+\mathcal{O}(r_0 ^{-1})=\mathcal{O}(r_0 ^{-1})$.

To simplify our next set of calculations, let us write for every $r \in I_{0}$, $r=r_0 +u$ with $u \in[-\epsilon,\epsilon]$. Note that
\begin{equation}\label{eq:powerbd}
    (r_0 + u)^{-p} = r_0 ^{-p} \left(1+ \frac{u}{r_0} \right)^{-p} = r_0 ^{-p} \left[1-\frac{pu}{r_0} + \mathcal{O}(r_0 ^{-2}) \right]
\end{equation}
Also, we can similarly bound the logarithmic term $(\log r)^{\beta}$ as follows.
\begin{equation*}
\begin{split}
    \log(r_0 + u) & = \log r_0 + \log \left(1+\frac{u}{r_0} \right) \\
    & = \log r_0 + \frac{u}{r_0} + \mathcal{O}(r_0 ^{-2}) 
\end{split}    
\end{equation*}
Hence, \begin{equation}\label{eq:logbd}
    \begin{split}
        (\log r)^{\beta} & = (\log r_0)^{\beta} \left[1+\frac{u}{r_0 \log r_0} + \mathcal{O}\left(\frac{1}{r_o ^2}\right)\right] \\
        & = (\log r_0)^{\beta}[1+\mathcal{O}((\log r_0)^{-1})]
    \end{split}
\end{equation}
Multiplying the orders of \ref{eq:powerbd}, and \ref{eq:logbd}, we have \[
k(r)\pi_{\text{poly},\beta}(r) = C r_{0} ^{-p} (\log r_0)^{\beta} \left[1+\mathcal{O}((\log r_0)^{-1})+\mathcal{O}(r_0 ^{-1})\right]
\]
Integrating over $I_0$, we have 
\begin{align*}
    A_1 & = \int_{r_0 - \epsilon}^{r_0 + \epsilon} k(r) \pi_{\text{poly},\beta}(r) dr \\
    & = C r_0 ^{-p}(\log r_0)^{\beta} \left[1+\mathcal{O}((\log r_0)^{-1}) \right] \int_{-\epsilon}^{\epsilon} du \\
    & = 2\epsilon C r_0 ^{-p} (\log r_0 )^{\beta}\left[1+\mathcal{O}((\log r_0)^{-1}) \right] \\
    & = C' (\log r_0)^{\beta} r_0 ^{-p} [1+o(1)]
\end{align*}
where we rename $C' \coloneqq 2 \epsilon C$ and absorb the $\mathcal{O}((\log r_0)^{-1})$ term into $o(1)$. The error term $o(1)$ holds uniformly in $\beta$ on any compact subset of $(0,\infty).$ We can lower bound $A_1 + A_2$ as follows: the integral $\int_{0}^{\eta r_0} r^{-p}(\log r)^{\beta} dr$ converges to a finite constant $C_1 \coloneqq (p-1)^{-(\beta+1)}\Gamma (\beta+1)$ and $k(r) \ge \exp(-A \omega^2) \rightarrow 1$. Looking at the tail $(\eta r_0,\infty)$,
\begin{align*}
    \int_{\eta r_0}^{\infty} \pi_{\text{poly},\beta}(r) k(r) dr & = r_0 ^{-(p-1)}(\log r_0)^{\beta} \int_{\eta}^{\infty} x^{-p} [1+\mathcal{O}(r_0 ^{-2})]\left(1+\frac{\log x}{\log r_0} \right)^{\beta} dx \\
    & = C_2 r_0 ^{-(p-1)} (\log r_0)^{\beta}[1+o(1)]
\end{align*}
Because $p>1$, the constant head $(0,\eta r_0)$ dominates and thus, $A_1 + A_2 = C_1[1+o(1)]$.
Then, the ratio \[\frac{A_1}{A_1 +A_2} = \frac{C'(\log r_0)^{\beta} r_0 ^{-p}}{C_1 [1+o(1)]} = \Theta(r_0 ^{-p} (\log r_0)^{\beta}) \]
So, the posterior risk is
\[
R_{\text{poly},\beta,\epsilon}(r_0) = 1-\Theta(r_0 ^{-p} (\log r_0)^{\beta}).
\]
\end{proof}
\subsection{When the parallax error $\sigma_\omega^2$ is unknown}\label{sec:unknownsigma}
We now treat $\sigma_\omega^2$ as an \emph{unknown scalar} (not a known function of~$\omega$) and place a prior on it. The joint posterior is
\[
\pi(r,\sigma_\omega^2\mid \omega)\ \propto\ 
(\sigma_\omega^2)^{-1/2}\exp\!\Big(-\tfrac{(\omega-1/r)^2}{2\sigma_\omega^2}\Big)\,
\pi(r)\,p(\sigma_\omega^2).
\]
Let $\sigma_\omega^2\sim \mathrm{IG}(\alpha,\beta)$ with density
$p(\sigma^2)\propto (\sigma^2)^{-\alpha-1}\exp(-\beta/\sigma^2)$, $\alpha,\beta>0$. Marginalizing $\sigma_\omega^2$ gives the Student–$t$ kernel in the $t\coloneqq r^{-1}$ - space:
\[
p(\omega\mid t)\ \propto\ \Big(\beta+\tfrac{(\omega-t)^2}{2}\Big)^{-(\alpha+\frac12)}
\ \equiv\ \Big(1+\tfrac{(\omega-t)^2}{\nu s^2}\Big)^{-\frac{\nu+1}{2}},
\]
with degrees of freedom $\nu=2\alpha$ and scale $s^2=\beta/\alpha$. Hence the posterior for $t$ is
\[
\pi(t\mid \omega)\ \propto\ \pi_t(t)\,\Big(\beta+\tfrac{(\omega-t)^2}{2}\Big)^{-(\alpha+\frac12)},
\qquad \pi_t(t)=\pi(r=1/t)/t^2.
\]
Consequently, the marginal posterior for $r$ is
\[
\pi(r\mid \omega)\ \propto\ \pi(r)\,\Big(\beta+\tfrac{(\omega-1/r)^2}{2}\Big)^{-(\alpha+\frac12)}.
\]
If $\pi(r)\sim C\,r^{-2}$ as $r\to\infty$ (e.g., Reciprocal–Gaussian, Half–Cauchy) then $(\omega-1/r)^2\to \omega^2$ and
\[
\pi(r\mid \omega)\ \asymp\ C'\,r^{-2},\qquad r\to\infty.
\]
If $\pi(r)\sim C\,(\log r)^{\beta}/r^2$ (Product Half–Cauchy type), then
\[
\pi(r\mid \omega)\ \asymp\ C'\,\frac{(\log r)^{\beta}}{r^2},\qquad r\to\infty.
\]
Thus, with a proper $\mathrm{IG}(\alpha,\beta)$ prior on $\sigma_\omega^2$, the posterior tail in $r$ matches the prior tail order; heavy (polynomial or log–polynomial) tails in $\pi(r)$ persist after marginalizing the noise variance.

\subsection{Convergence Diagnostics}
To assess the convergence and efficiency of the MCMC sampling process for different prior distributions in our simulation setting (Section \ref{sec:simuls}), we calculated the mean R-hat \citep{gelmanrhat} and mean Effective Sample Size (ESS) \citep{practicalMCMC} for each prior. The R-hat statistic measures convergence, with values close to 1.0 indicating good mixing of the chains. The ESS quantifies the effective number of independent samples, with higher values suggesting more efficient sampling. 

Table~\ref{tab:mcmc_diagnostics} summarizes the MCMC diagnostics for the models using different priors.
\begin{table}[H]
    \centering
    \caption{MCMC Diagnostics for Different Priors}
    \label{tab:mcmc_diagnostics}
    \footnotesize{
    \begin{tabular}{lrr}
        \toprule
        \textbf{Prior} & \textbf{Mean ESS} & \textbf{Mean R-hat} \\
        \midrule
        Half-Cauchy            & 4976.061 & 0.9998 \\
        Gamma             & 6818.535 & 0.9997 \\
        Product Half-Cauchy & 3000.933 & 1.0010 \\
        Weibull           & 6630.414 & 0.9998 \\
        Reciprocal-Gaussian                & 5551.344 & 0.9998 \\
        Inverse-Gamma        & 5953.725 & 0.9998 \\
        \bottomrule
    \end{tabular}}
\end{table} 
The mean R-hat values across all models are close to 1.0, indicating good convergence of the MCMC chains. However, the Product Half-Cauchy prior has a lower mean ESS compared to the other priors, suggesting that while the chain has converged, the sampling efficiency is reduced. This could be due to higher auto-correlation in the samples, which warrants further investigation.

\subsection{Posterior Predictive Checks}\label{sec:ppc}
Posterior predictive checks (PPCs) are used to assess the fit of the models by comparing the observed data with data simulated from the posterior distribution. The goal is to evaluate whether the model is capable of generating data that looks similar to the observed data.

\begin{figure}[!ht]
    \centering
    \includegraphics[width=0.7\linewidth]{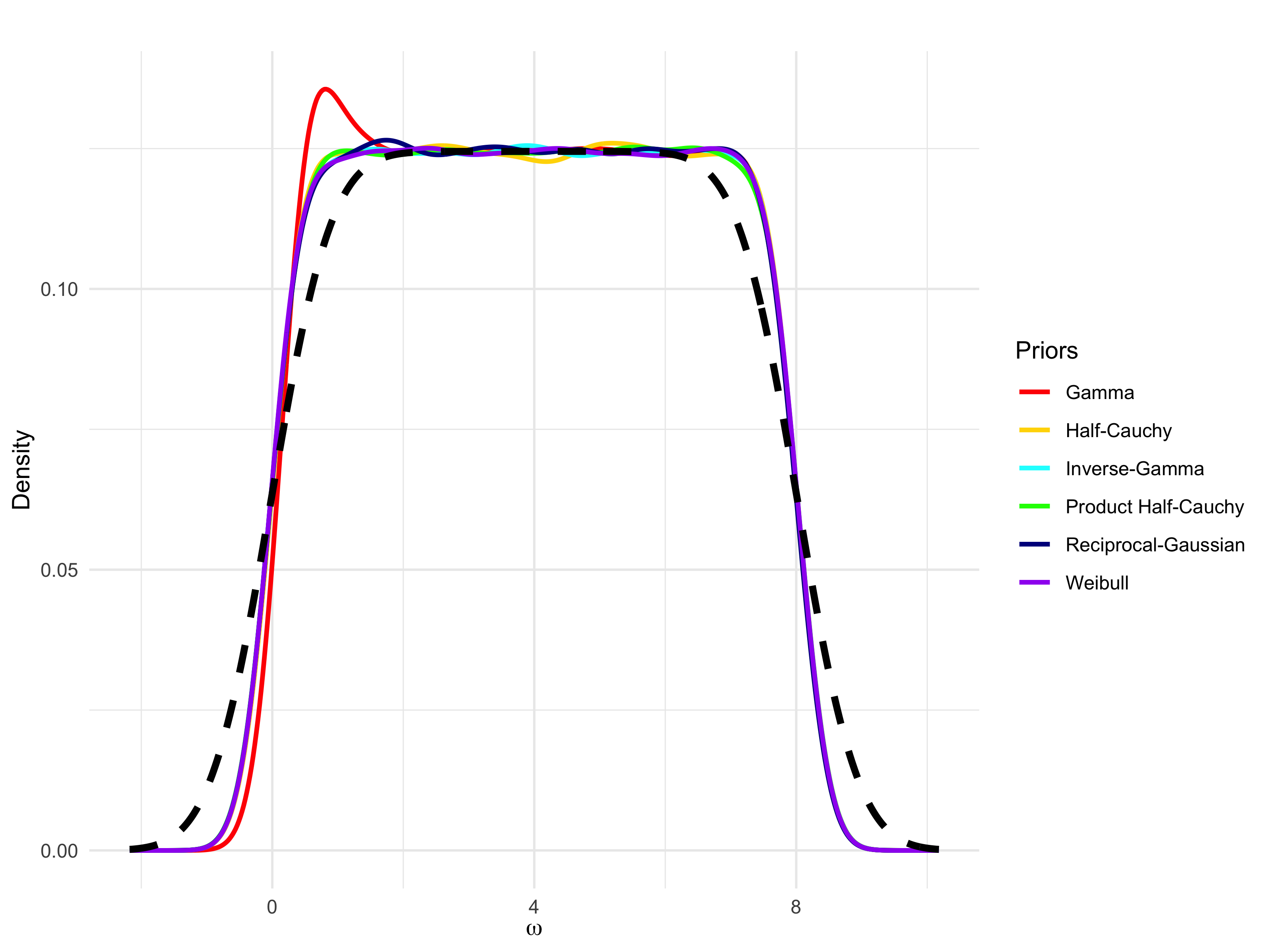}
    \caption{\footnotesize{Posterior predictive densities for different priors}}
    \label{fig:posterior_pred_checks}
\end{figure}

Figure~\ref{fig:posterior_pred_checks} shows the posterior predictive densities for models using different prior distributions, overlaid on the observed data (black dashed line). Each colored line represents the predictive density generated by a model using a different prior.
The plot demonstrates that most of the priors result in posterior predictive distributions that closely follow the observed data, as indicated by the overlap between the colored lines and the black dashed line. The Gamma prior, represented by the red line, shows a slight deviation at lower values, which might indicate that this model is somewhat over-fitting or under-fitting in this region. The other priors (Half-Cauchy, Inverse-Gamma, Product Half-Cauchy, Reciprocal-Gaussian, and Weibull) show predictive densities that are very similar to each other and closely match the observed data, suggesting that these models fit the data well.

\subsection{Sensitivity Analysis with Respect to Measurement Error}\label{sec:sensitivity}
In practice, the precision of parallax measurements depends critically on the number of photons received by the telescope and other observational conditions \citep{mcgraw08}. While missions such as \emph{Hipparcos} and \emph{Gaia} are designed to achieve very low measurement error levels (i.e., small $\sigma_{\omega}$), it is instructive to explore how our inference procedure performs when $\sigma_{\omega}$ is varied. In particular, we assess the robustness of our candidate priors under increasing levels of measurement noise by evaluating the squared error. 
\begin{figure}[H]
  \centering
  \includegraphics[width=0.9\linewidth]{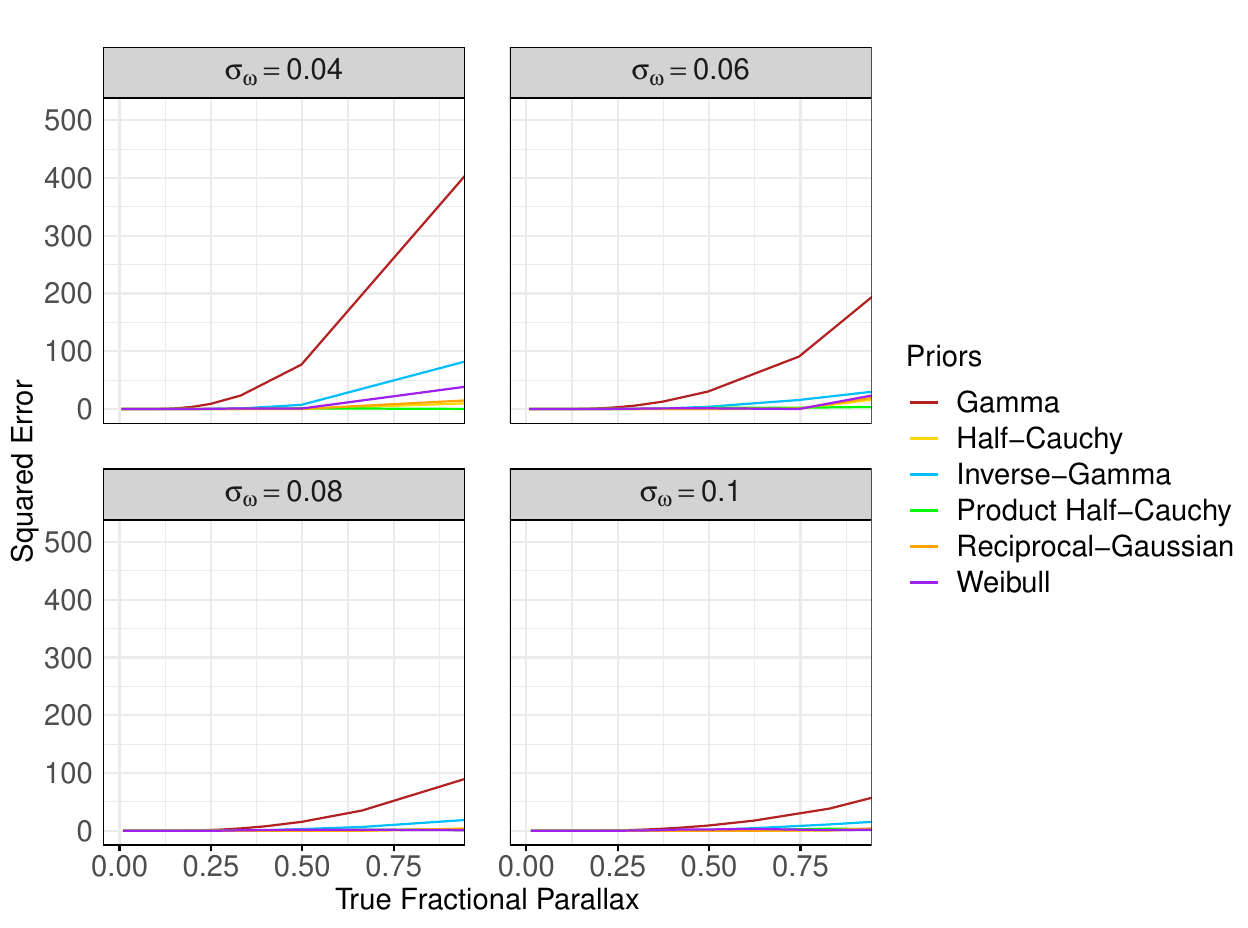}
  \caption{\footnotesize{Sensitivity of squared error as a function of the true fractional parallax error for various choices of $\sigma_{\omega}$.}}
  \label{fig:sensitivity}
\end{figure}
At first glance, one might expect that an increase in $\sigma_{\omega}$, indicating larger measurement error, would result in higher squared error. However, our simulation experiments (see Figure~\ref{fig:sensitivity}) reveal that the squared error tends to decrease as $\sigma_{\omega}$ increases. An important factor contributing to this can be explained by the fact that the relationship between parallax and distance is nonlinear; small errors in $\omega$ are amplified when computing $r = 1/\omega$, particularly when $\omega$ is close to zero. Increasing $\sigma_{\omega}$ effectively smooths the effect of this nonlinear transformation. A less informative likelihood prevents extreme estimates caused by the inversion, which results in a more stable estimation process and hence lower squared errors.

In settings where the observational noise is relatively large, appropriately chosen heavy-tailed priors can mitigate the adverse effects of the nonlinearity in the distance inversion, yielding more robust distance estimates. This phenomenon is particularly relevant for observations of faint or distant stars, where the measurement errors are inherently larger.

\bibliographystyle{ba}
\bibliography{paper-ref,dattaref,horseshoe-plus,refs}

\begin{acknowledgement}
Dr. Datta gratefully acknowledges support from the National Science Foundation (NSF CAREER Award DMS-2443282).
\end{acknowledgement}

\end{document}

%% file: math-commands.tex
\renewcommand{\L}{{\mathcal L}}

\newcommand{\ben}{\begin{enumerate}}
\newcommand{\een}{\end{enumerate}}
\newcommand{\beq}{\begin{equation}}
\newcommand{\eeq}{\end{equation}}
\newcommand{\bde}{\begin{description}}
\newcommand{\ede}{\end{description}}

\newcommand{\NormRV}{\mathcal{N}}

\newcommand{\UnifRV}{\mathcal{U}}

\usepackage{booktabs,array}

\newcount\rowc